\documentclass{ametsocV6.1}

\title{Trends in Northern Hemispheric Snow Presence}

\authors{Yisu Jia,\aff{a}\correspondingauthor{Yisu Jia, y.jia@unf.edu} 
Robert Lund,\aff{b} 
Jiajie Kong,\aff{b} 
Jamie Dyer,\aff{c}
Jonathan Woody,\aff{d}
J. S. Marron,\aff{e}
}

\affiliation{\aff{a}{Department of Mathematics and Statistics,\\ University of North Florida}\\
\aff{b}{Department of Statistics, \\
The University of California, Santa Cruz}\\
\aff{c}{Department of Geosciences,\\Mississippi State University}\\
\aff{d}{Department of Mathematics and Statistics,\\Mississippi State University}\\
\aff{e}{Department of Statistics and Operations Research,\\University of North Carolina at Chapel Hill}
}

\abstract{This paper develops a mathematical model and statistical methods to quantify trends in presence/absence observations of snow cover (not depths) and applies these in an analysis of Northern Hemispheric observations extracted from satellite flyovers during 1967-2021.  A two-state Markov chain model with periodic dynamics is introduced to analyze changes in the data in a grid by grid fashion.  Trends, converted to the number of weeks of snow cover lost/gained per century, are estimated for each study grid.  Uncertainty margins for these trends are developed from the model and used to assess the significance of the trend estimates.  Grids with questionable data quality are identified.  Among trustworthy grids, snow presence is seen to be declining in almost twice as many grids as it is advancing.  While Arctic and southern latitude snow presence is found to be rapidly receding, other locations, such as Eastern Canada, are experiencing advancing snow cover.}

\begin{document}

\maketitle

%
%
%
\noindent \textbf{Disclaimer:}\textit{
This Work has not yet been peer-reviewed and is provided by the contributing Author(s) as a means to ensure timely dissemination of scholarly and technical Work on a noncommercial basis. Copyright and all rights therein are maintained by the Author(s) or by other copyright owners. It is understood that all persons copying this information will adhere to the terms and constraints invoked by each Author's copyright. This Work may not be reposted without explicit permission of the copyright owner.}

\noindent \textbf{Copyright notice:}
\textit{This Work has been submitted to Journal of Climate. Copyright in this Work may be transferred without further notice.}
\statement The purpose of this project is to quantify how the Northern Hemisphere’s snow cover has changed. Snow cover plays a critical role in the global energy balance due to its high albedo and insulating characteristics and is therefore a prominent indicator of climate change. On a regional scale, the spatial consistency of snow cover influences surface temperatures via variations in absorbed solar radiation, while continental-scale snow cover acts to maintain thermal stability in Arctic and subarctic regions, leading to spatial and temporal impacts on global circulation patterns. Changing snow presence in Arctic regions will influence large scale releases of carbon and methane gas. Given the importance of snow cover, understanding its trends aids our understanding of climate change.

%
%
%

%

\section{Introduction}
Snow cover plays a critical role in the Earth's hydrological processes and its impact on the broader global climate is of great interest \citep{barnett2005potential, karl2009global, goudie2018human, van2009widespread}.  Snow greatly influences the global energy balance due to its high albedo and insulating characteristics and is therefore a prominent indicator of climate change \citep{liston2011changing, mote2003trends, lawrence2010contribution, callaghan2011changing}.  On a regional scale, the spatial consistency (patchiness) of snow cover can influence surface temperatures via horizontal variations in absorbed solar radiation.   Continental-scale snow cover acts to maintain thermal stability in the Arctic and subarctic regions, possibly inducing changes in global circulation patterns attributable to large-scale releases of carbon and methane gas \citep{zona2016cold}.  While the amount of water available in the snowpack is quantified in snow depths and/or snow water equivalents (SWE), areal snow presence/coverage defined by snow cover extent (SCE) is useful for estimating the location and availability of regional water resources \citep{mote2018dramatic, serreze2000observational, robinson1993global}.

Remotely sensed satellite images are common sources of SCE data; these provide spatial and temporal observations that can be used in regional and continental-scale analyses. Satellite data is used here to estimate SCE trends, allowing us to assess SCE changes over time and space.

Satellite-derived SCE data come in a binary format, with snow presence being recorded as unity and snow free ground being assigned zero.  Some mid-latitude locations have sporadic snow coverage, with snow cover typically lasting only a few weeks at a time, even during the height of winter.  The majority of our work lies with introducing a mathematical model and developing the statistical methods needed to analyze trends in autocorrelated and binary-valued sequences, yet flexible enough to adapt to the data from our many study pixels (grids).  

Statistical analysis of snow data has been debated in the climate literature, especially in regard to trend and uncertainty assessment --- see \citep{yue2002influence} and the references therein.  Here, a flexible mathematical model and rigorous accompanying statistical methods are used to assess trends and accurately assess their uncertainity margins. Some nuances arise in this pursuit.  First, as our SCE data are recorded weekly, annual periodicity needs to be taken into account.  Second, since SCE data is correlated, with snow presence in a week making snow presence in adjacent weeks more likely, serial autocorrelation should also taken into account in trend uncertainity quantification.  Finally, previous authors have noted data quality issues \citep{bormann2018estimating, estilow2015long} in some grids that need to be addressed, without pinpointing the specific problematic grids.  We carefully address this issue below.  The general pattern of results found here agrees with trends found in other studies using more rudimentary statistical approaches \citep{brown2011northern, lemke2007, notarnicola2022overall}.

The rest of this paper proceeds as follows.  Section 2 describes the SCE data used in this study and its nuances. Section 3 introduces the mathematical model and statistical methods needed to quantify the problem, including the all-important uncertainties in the trend margins.  Section 4 presents a simulation study, showing that model parameters can be accurately estimated from our binary data from a half century of weekly data.  Section 5 presents two case studies, analyzing observations from a grid in North Dakota that is actually experiencing increasing snow cover.  We also give an example of data from a grid with poor data quality.  Section 6 presents results for the entire Northern Hemisphere (NH) and discusses our general findings and their implications.  Section 7 concludes with comments and remarks. 

\section{Data}

\begin{figure}[ht]
\centering
\includegraphics[width=1\textwidth]{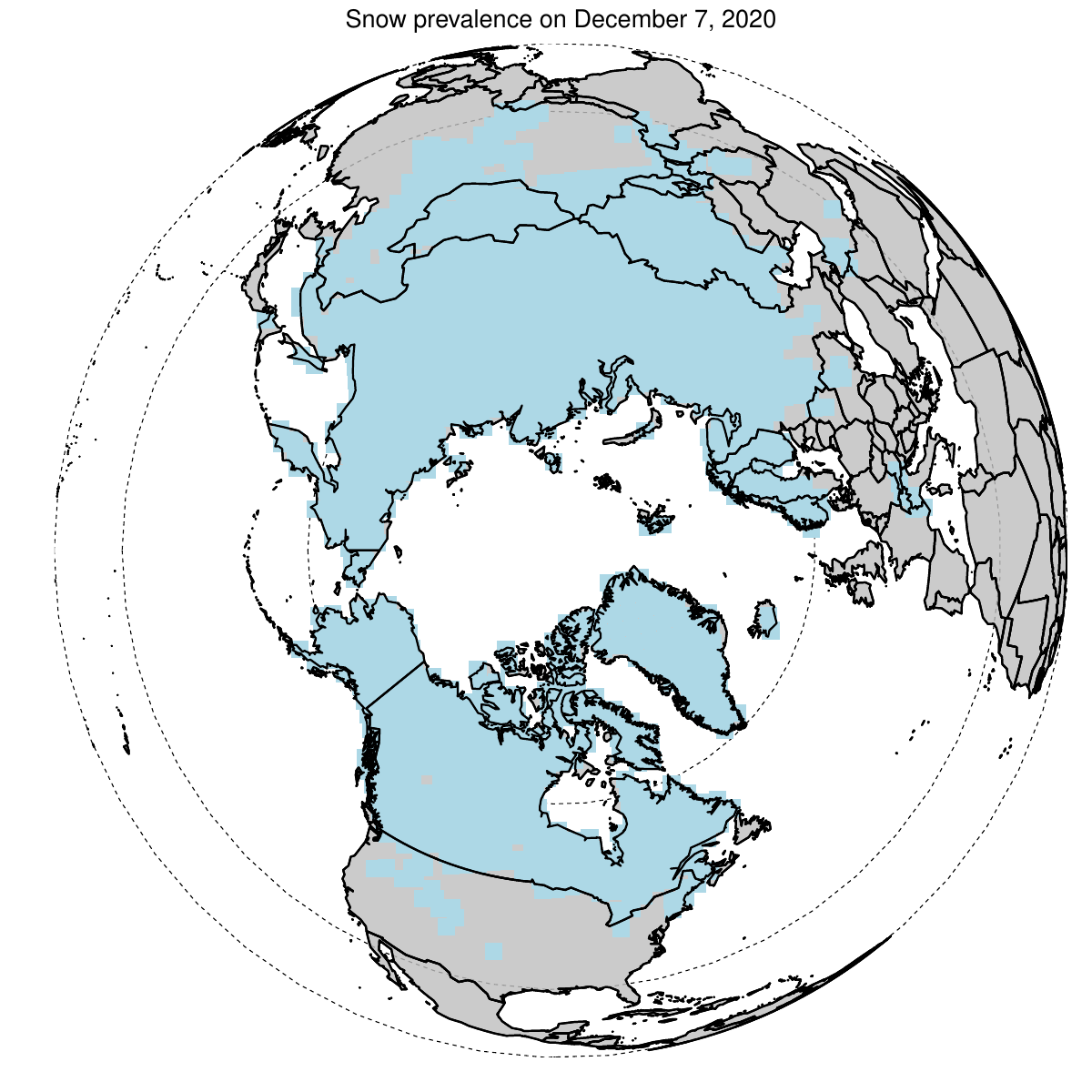}\\
\caption{NH snow coverage reported for the first week of December, 2020.}
\label{fig: sample week}
\end{figure}

Our data were collected from weekly satellite flyovers, with SCE values being estimated manually by meteorologists for each grid.  This study uses data from the Climate Data Record as developed by the National Oceanic and Atmospheric Administration (NOAA), which is housed at the Rutgers University Snow Lab at \url{http://climate.rutgers.edu/snowcover/}. This study examines the January 1967 - July 2021 period.  For grid structure, the data use NOAA's $89 \times 89$ Cartesian grid that overlies a polar stearographic projection of the NH.  The SCE during the first week in December 2020 is plotted in Figure \ref{fig: sample week} for feel.

Thorough descriptions of the data are provided in \citet{dye2002variability} and \citet{estilow2015long}. Production of the data is discussed in \citet{wiesnet1987discussion} and \citet{robinson1993global}.  Before June of 1999, NOAA used the first clear-sky day during each week to estimate the SCE.  If the grid contains at least 50 percent snow coverage, its SCE was assigned unity; otherwise, it is assigned zero. 

With the introduction of the Interactive Multisensor Snow and Ice Mapping System, the methods used to estimate SCE changed in June 1999.  These methods use different data and a refined partition of the NH; changes are detailed in \citep{estilow2015long}. \citet{brown2007assessment} did not find evidence of inhomogenities over Northern Canada before and after the 1999 change; however, \citet{dery2007recent} claim that pre-1999 methods overestimate snow presence in mountainous regions during Spring ablation. An analysis of the 1999 change is provided here later.   

\begin{figure}[ht]
\centering
\includegraphics[width=0.8\textwidth]{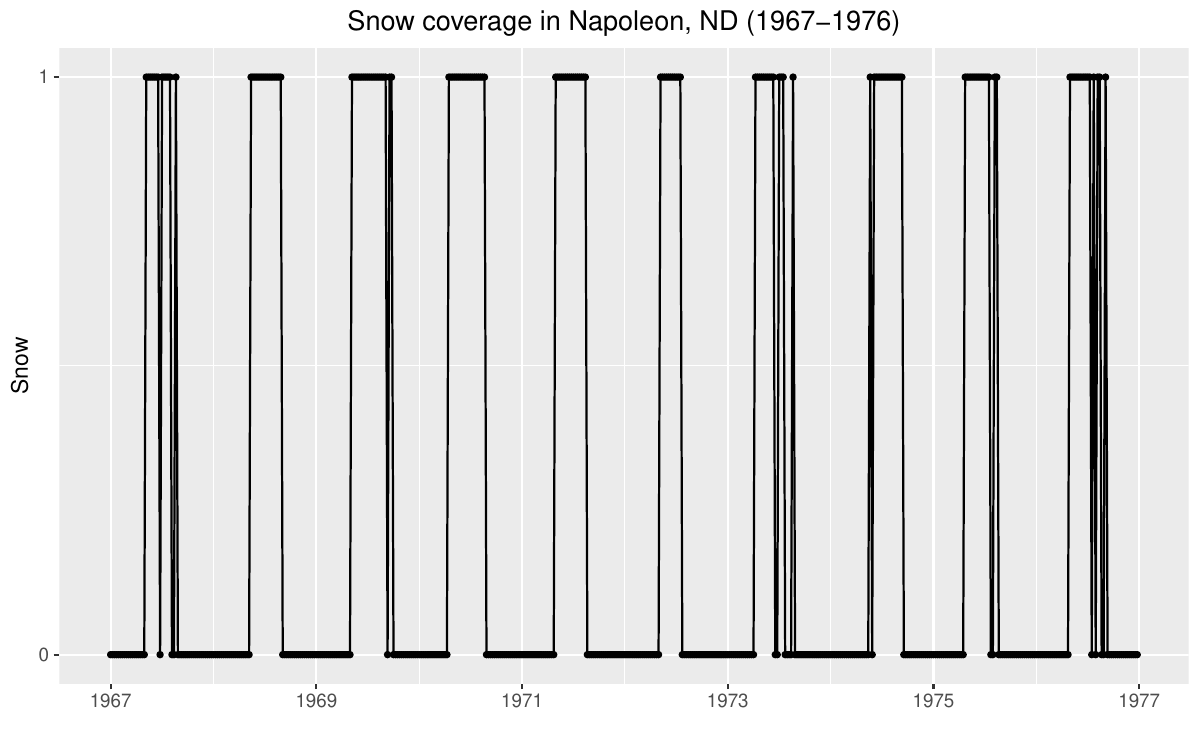}\\
\caption{Ten years of snow presence/absence (1967-1976) for the Napoleon, ND grid.}
\label{napoleon}
\end{figure}

Figure \ref{napoleon} displays ten years of observations for a grid located near Napoleon, ND, from 1967-1976.  This grid will be analyzed in detail in Section 5.  The graph reveals the ephemeral nature of snow processes here, starting each year circa November and typically lasting through early April.  Once snow cover is present, it usually stays through Spring ablation; however, winters exist where snow cover oscillates (1967-1968 and 1973-1974 for example).

Our study partitions the NH into 3,011 grids over land. Winter centered years are used here so that the first week of any year corresponds to the first week of August. This scaling prevents a single winter's snow record from lying within two distinct years.  Shifting in this manner is done for convenience only --- the scaling does not influence any trends.  Each grid was categorized into four subgroups, depending on its data. Grid Group 1 includes all grids that reported 10 or less weeks of snow cover during the 1967-2020 period of record (2,808 weeks).  This group also contains any grid that reported 10 or fewer weeks of bare ground over the record period.  Group 1 grids primarily lie in the southerly latitudes of the NH, which rarely experience snow, or the interior Greenland icecap, which is almost always under snow cover. All 1,131 Group 1 grids were excluded as any trends computed from these records lack sufficient information/variability for any meaningful statistical analysis.  Group 2 contains 72 grids that were insufficiently fitted by our model (our model is explained in the next section). While these grids all had more than 10 snow/bare ground weeks during the 2,808 week study period, they typically did not have many more.  These grids were primarily located in Southern China, the Southern United States, and Coastal Greenland.  Group 3 contains 195 grids where the data record appears untrustable, especially pre-1999. These grids all reside in mountainous regions of the NH (Rockies, Alps, Caucasus, Scandinavia, and Himalayas) and are omitted from further analysis.  Figure \ref{fig:bad_grid_sample} plots the data from an example Group 3 grid located in the Chinese Himalayan Mountain Range.   Several issues are apparent.  The top plot shows that some of the earlier years in the record have no snow cover in winter weeks, but some snow cover during summer weeks.   The bottom plot reveals that the pre-1999 years report very little snow cover compared to the post-1999 years.  While the methodological revisions in 1999 may render the post-1999 data believable, this grid is best excluded in a trend analysis.  

\begin{figure}[ht]
\centering
\includegraphics[width=1\textwidth]{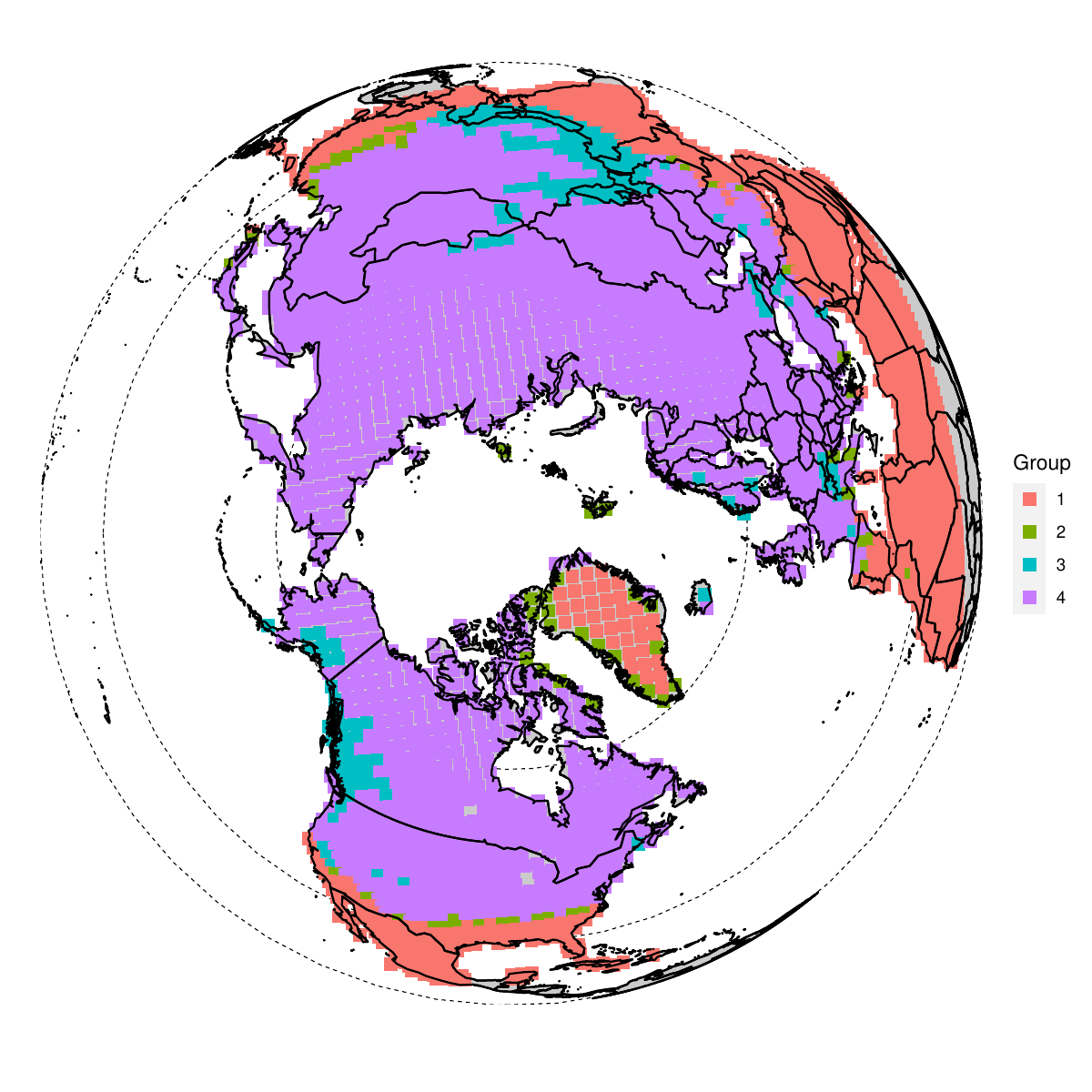}\\
\caption{A graphical partition of the grids in the study.  The violet colored grids (Group 4) were deemed analyzable in this study.  The grids in Group 1 are excluded because they either lie in icecaps or tropical localities and have little variability. Group 3 grids were excluded as their data were deemed unreliable. Group 2 contains a small number grids that are too poorly described by our model (for various reasons) to report an analysis.}
\label{fig:group}
\end{figure}

\begin{figure}[ht]
\centering
\includegraphics[width=0.9\textwidth]{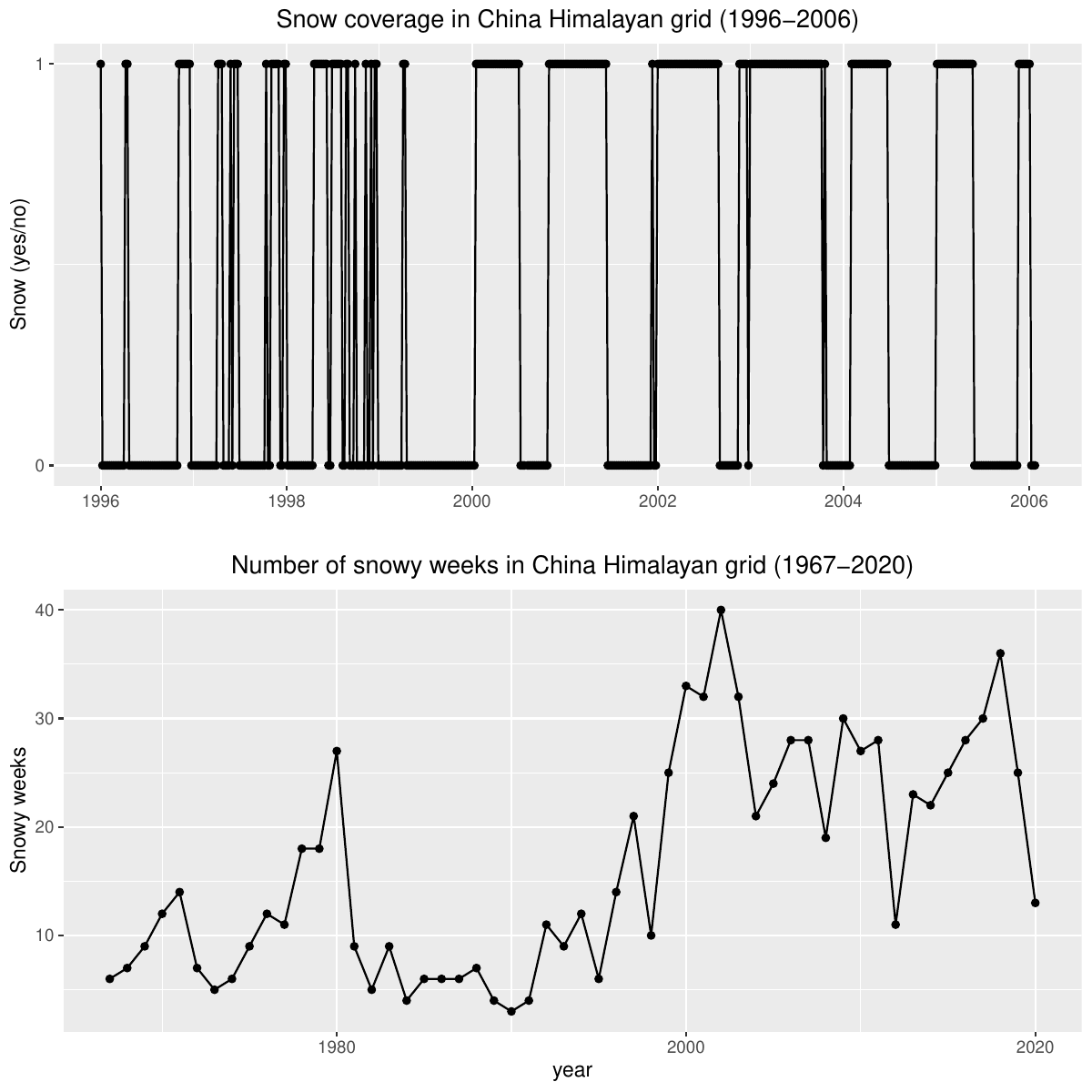}\\
\caption{A grid from the Himalayas with untrustable data. Top: Ten years of snow presence/absence from 1996-2006; Bottom: The number of snow covered weeks during the 1967-2020 period.}
\label{fig:bad_grid_sample}
\end{figure}

The other 1,613 grids were placed in Group 4 and will be further analyzed.  These grids were examined on a one-by-one basis and were deemed to produce a reliable model fit (this is done through a variety of diagnostic procedures on the gradient step and search likelihood maximization).   Figure \ref{fig:group} depicts the Group category of all grids; notice that the 1,613 violet grids where our model fit was deemed reliable cover most areas of the NH where snow is seasonally persistent.  A spreadsheet containing the group numbers of our grids is available from the authors upon request.

Several previous studies of this data exist.  \citet{dery2007recent} studies the data from January 1972 - December 2006.  \citet{dery2007recent} report significant temporal autocorrelation in the data, at both weekly and annual scales.  Autocorrelation makes some statistical methods such as Sen's slope troublesome for trend analysis as uncertainties cannot be accurately estimated with such a non-parametric method \citep{yue2002influence}.  Negative trends in SCE area are reported in \citet{dery2007recent} from March through June.  Figure 4.3 in \citet{lemke2007} shows March-April snow cover departures by subtracting the percentage coverage (by grid) of weeks with snow cover from 2004-1988 minus the same percentage coverage during 1967-1987.  While it is not clear how to interpret such a statistic as any type of smooth trend, the largest reductions in that study occurred roughly between the $0^{\circ}$C and $5^{\circ}$C isotherms.

\section{Model and Estimation}

\subsection{The Model}

Our methods use a two-state Markov chain model on the states $\{ 0,1 \}$ to describe the series for a fixed grid. State zero indicates lack of snow and state one signifies snow cover. The transition probability matrix of this chain from week $t-1$ to week $t$ is parameterized as
\[
\mathbf{P}(t)=
\left[ \begin{array}{cc}
p_{0,0}(t) & p_{0,1}(t) \\
p_{1,0}(t) & p_{1,1}(t)
\end{array}
\right].
\]
\noindent Here, $p_{0,1}(t)$ is the probability that snow cover is present at time $t$ given that it is absent at time $t-1$.  The other three elements in the matrix are similarly interpreted.  There are only two free quantities in $\mathbf{P}(t)$ at any $t$ since $p_{0,0}(t) = 1 - p_{0,1}(t)$ and $p_{1,0}(t) = 1 - p_{1,1}(t)$.

Let $\{ X_t \}$ denote the two-state snow presence/absence chain.  Then $X_t=1$ means that snow cover is present at time $t$ and $X_t=0$ means that snow is absent at time $t$.  The marginal distribution of $X_t$ at time $t$ will be denoted by $\boldsymbol{\pi}(t)=(\pi_0(t), \pi_1(t)) =(P(X_t=0), P(X_t=1))$.  Because the chain commences with an observation in August, the startup condition $\boldsymbol{\pi}(1)=(1,0)$ is taken, signifying that the chain starts with bare ground.  With this initial distribution, $\boldsymbol{\pi}(t)$ is computed via
\begin{equation}
\boldsymbol{\pi}(t) = \boldsymbol{\pi}(1) \prod_{k=2}^t {\mathbf P}(k).
\label{transition}
\end{equation}
For each pair of times $t_1 < t_2$ in $\{ 1, \ldots, N \}$, the transition matrix 
\[
{\bf P}^*(t_1,t_2)=\prod_{t=t_1+1}^{t_2}{\bf P}(t)
\]
contains the four transition probabilities of snow cover/absence from time $t_1$ to time $t_2$.

Since $p_{0,1}(t)$ and $p_{1,0}(t)$ are probabilities, they take values in $[0,1]$.  Hence, these quantities are modeled with the logistic-type link
\[
p_{0,1}(t) = \frac{1}{1+\exp(-m_{t})}, \quad
p_{1,0}(t) = \frac{1}{1+\exp(-m^*_{t})},
\]
\noindent where $m_t$ and $m^*_t$ contain seasonal effects and trend parameters.  These quantities are posited to have the additive form
\[
m_{t}   = \mu_{t}   + \alpha t,    \quad 
m^*_{t} = \mu^*_{t} + \alpha^* t,
\]
\noindent where the parameters are clarified as follows.  First, $T$ is the period of the data.  For the weekly observations analyzed here, the period $T=52$ weeks is forced to the data by omitting any observations that occur at the end of July (one day during non leap years and two days during leap years). This tactic results in little loss of precision.  The parameters $\mu_{t}$ and $\mu^*_{t}$ contain seasonal effects that are sinusoidaly parametrized as 
\[
\mu_{t} = A_0 +
A_1   \left[\cos \left(\frac{2\pi(t-\kappa)}{T}\right)\right],
\quad
\mu^*_{t} = A_0^* +
A_1^* \left[\cos \left(\frac{2\pi(t-\kappa^*)}{T}\right)\right].
\]
\noindent Observe that $\mu_t$ and $\mu_t^*$ are periodic with period $T=52$ weeks and obey $\mu_{t+T}= \mu_t$ and $\mu^*_{t+T}=\mu^*_t$. The quantities $A_0$ and $A_0^*$ govern the length of the snow season.  For example, when $A_0>0$, the season where snow is present tends to last longer than the snow free season (and vice versa). The parameters $A_1$ and $A_1^*$, which are assumed positive for mathematical identifiability of the cosine waves, control how fast snow to bare ground transitions take place (and vice versa).  The parameters $\tau$ and $\tau^*$ are phase shifts.  Since $p_{0,1}(t)$ and/or $p_{1,0}(t)$ are maximized when $m_t$ and/or $m_t^*$ is maximized, and the cosine function is maximized when its argument is zero, $p_{0,1}(t)$ is maximized at week $\kappa$, which is typically in the late Fall or early winter, and $p_{1,0}(t)$ is maximized at week $\kappa^*$, which typically occurs in the late winter or early spring. The parameters $\alpha$ and $\alpha^*$ are linear trend parameters and govern how fast snow cover changes are happening.  While the above model has a linear time trend and a simple cosine seasonal cycle, other forms of trends and seasonality could be considered if needed.

Our periodic Markov chain model allows $X_t$ to be autocorrelated in time $t$.  Indeed, week to week SCE data should be correlated:  if snow is present/absent at week $t$, it is more likely to be present/absent at week $t+1$.  Good models for snow depth processes are also correlated in time and have a Markov structure.  Indeed, \cite{Woody_WRR} argues for a Markov structured storage model for daily snow depths: the snow depth today is the snow depth yesterday, plus any new snowfall, minus any meltoff or compaction between yesterday and today.  Our model is also not a classical Probit count time series model as these are typically used for uncorrelated data; see \cite{Probit} for more on probit modeling.  A Markov model for binary data is parsimonious in that there are only two free parameters in $\mathbf{P}(t)$ for each fixed $t$.  While seasonal and trend features need to be incorporated into $\mathbf{P}(t)$ to handle the periodic nature of snow, the overall model is very parsimonious.  Comparing further, a time homogeneous Markov model for categorical sequences taking on $S$ distinct categories has $S(S-1)$ free parameters, which can be quite large for a large $S$.  Additional parameters would be needed to make this model periodic.

\begin{figure}[ht]
\centering
\includegraphics[width=\textwidth]{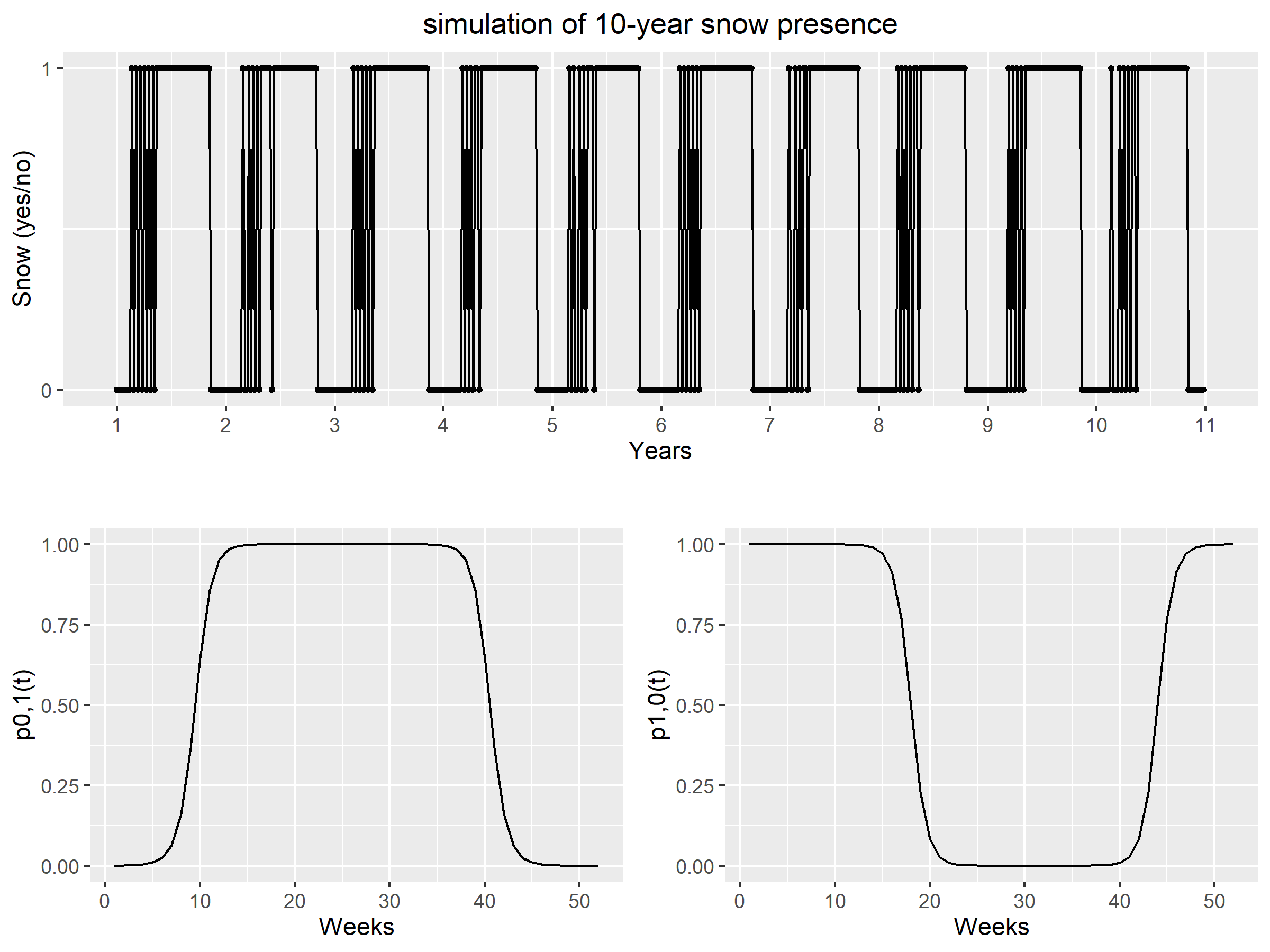}
\\
\caption{The parameters for $p_{0,1}(t)$ and $p_{1,0}(t)$ are $A_0=3, A_1=10, \tau=25, \alpha=0$, and $A_0^*=0, A_1^*=10, \tau^*=5, \alpha^*=0$ (no trend).}
\label{Fig1}
\end{figure}

Figure \ref{Fig1} shows a simulation of ten years of a binary snow presence process. The parameters chosen for $p_{0,1}(t)$ are $A_0=3, A_1=10, \tau=25, \alpha=0$, and those for $p_{1,0}(t)$ are $A_0^*=0, A_1^*=10, \tau^*=5, \alpha^*=0$; specifically, there is no trend in the simulated data.  One sees that each and every year, snow presence begins in the Fall and stays on the ground until Spring.  Oscillations between snow presence and bare ground occur in the Fall, and snow vanishes completely during the summer.  Additional simulations show that this simple Markov chain model produces a flexible suite of snow presence/absence series.

\subsection{Parameter Estimation}

Suppose that the binary data sample ${\mathbf X}=(X_1, \ldots, X_N)^\prime$ is available for a fixed grid.  We assume that $N$ is a multiple of $T$ to avoid trite work with fractional portion of years; this said, the methods are easily modified to accommodate fractional parts of years if needed. Let $d=N/T$ denote the total number of years of observations and work with observations labeled as the years $1, 2, \ldots , d-1$

Let $\mathbf{\Theta}$ denote all model parameters contained in $m_t$ and $m_t^*$.  These include $A_0, A_1, \kappa, \alpha$ and their starred counterparts.  The statistical likelihood of $\mathbf{\Theta}$, denoted by $L(\mathbf{\Theta}|{\mathbf X})$, can be derived from the Markov property and is
\begin{equation}
\ln(L(\mathbf{\Theta}|\mathbf{X})) =
\sum_{t=2}^N \ln \left( p_{X_{t-1}, X_t}(t) \right).
\label{eq: likelihood}
\end{equation}
The quantities $p_{i,j}(t)$ depend on $\boldsymbol{\Theta}$.  Numerically maximizing this likelihood provides estimates of the components in $\mathbf{\Theta}$, which will later be useful in assessing variability (uncertainty) margins of the trends.  The data $X_1, \ldots , X_N$ is held fixed in this maximization.  While explicit forms for the parameter estimators in $\boldsymbol{\Theta}$ do not exist, likelihood estimates can obtained by numerically maximizing the likelihood.  Likelihood maximization is a reasonably standard and stable numerical procedure, executable via many gradient step and search optimization routines.

\subsection{Trend Estimation and their Uncertainties}

Trends will be phrased in the number of snow weeks lost/gained per time. For example, future trends will be phrased as a loss of one week of annual snow cover over a century.  Quantifying this, let $S_n$ be the number of weeks of snow on the ground during year $n$:
\[
S_n = \sum_{\nu=1}^T 1_{[X_{(n-1)T+\nu}=1]},
\]
\noindent where $1_A$ denotes the indicator of the event $A$.  We make our trend inferences from the quantities $\mathbf{S} = (S_1, S_2, \ldots, S_d)^\prime$ ($\prime$ denotes matrix transpose).   The linear rate of SCE change is quantified by $\hat{\beta}$ defined by 
\begin{equation}
\label{trend}
\hat{\beta}=
\frac{\sum_{k=1}^d S_k(k - \bar{k})}
{\sum_{k=1}^d
(k - \bar{k})^2} =
\frac{\sum_{k=1}^d S_k(k - \bar{k})}{Q},
\end{equation}
where $\bar{k}=(d+1)/2$ is the average time index and the denominator can be verified as $Q=d(d+1)(d-1)/12$. While the units of $\beta$ are weeks of snow cover gained/lost per year, we will scale $\hat{\beta}$ to weeks of snow cover gained/lost per century for interpretability; this simply multiplies raw trends and their standard errors by 100.

Our next objective is to obtain a standard error for $\hat{\beta}$. Taking a variance in (\ref{trend}) gives
\[
\mbox{Var}(\hat{\beta})=\frac{ \sum_{k=1}^d \sum_{\ell=1}^d (k-\bar{k})(\ell-\bar{k}) \mbox{Cov}(S_k, S_\ell)}{Q^2}.
\]
This computation requires $\mbox{Cov}(S_n, S_{n+h})$ for every $h > 0$ and $n$ in $\{ 1, \ldots, d-h \}$.  Details for this computation are provided in the Appendix.  The standard error of $\hat{\beta}$ accounts for correlation aspects in the SCE data.

To statistically test whether or not SCE is changing, we want to test the null hypothesis that $\beta = 0$ against the alternative that $\beta \neq 0$.  Invoking asymptotic normality of the estimator $\hat{\beta}$, this is assessed through the $Z$-score
\[
Z = \frac{\hat{\beta}}{\mbox{Var}(\hat{\beta})^{1/2} },
\]
which is compared to the standard normal distribution to make conclusions.  One typically reports a $p$-value for the test to assess significance of the trends; this is illustrated in Section 6.

\section{A Simulation Study}

This section studies our model and estimation procedure via simulation, illustrating the model's capabilities and how parameters are estimated.

\begin{figure}[ht]
\centering
\includegraphics[width=\textwidth,height=8in]{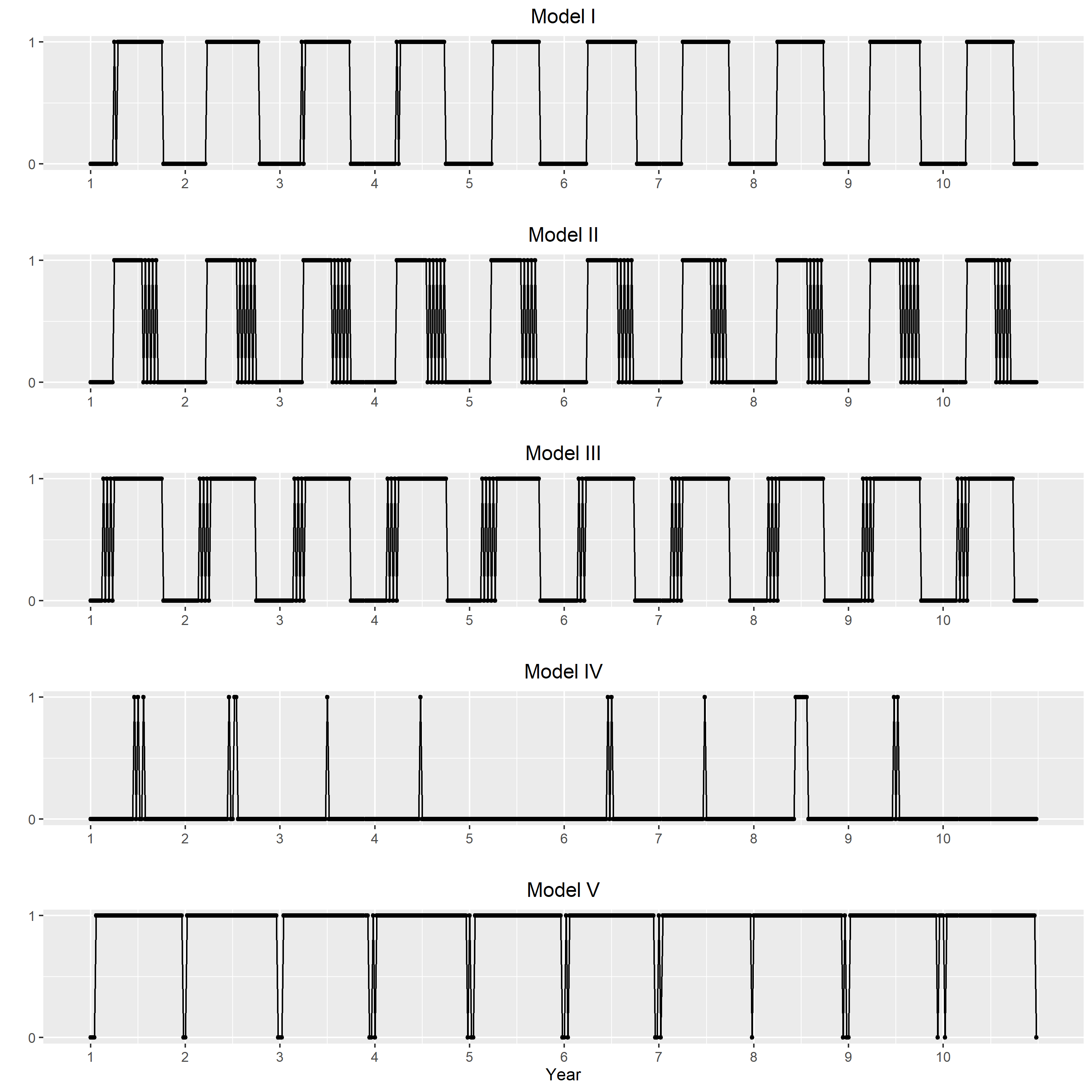}\\
\caption{Ten year sample series generated from Models I-V.}
\label{sim}
\end{figure}

To demonstrate the model's flexibility, Figure \ref{sim} provides ten year sample plots of snow presence/absence series generated by models for five sets of parameter values.   Only ten years of data are shown --- it becomes visually difficult to see data features with longer series (the plot becomes ``compressed").  Table 1 lists the parameter values that will be considered.   The unstarred parameters govern $p_{0,1}(t)$, which controls transitions from bare ground to snow cover; the starred parameters govern $p_{1,0}(t)$, which controls transitions from snow cover to bare ground.

\begin{table}[ht]
\label{modelspecs}
\caption{Sample Simulated Series}
\begin{center}
\begin{tabular}{ccccccccc}
Model & $A_0$ & $A_1$ & $\kappa$ & $\alpha$ & $A_0^*$ & $A_1^*$ & $\kappa^*$ & $\alpha^*$ \\
\hline
I  &  0  & 30 & 25  & 0 & 0   & 30  & 0  & 0  \\
II &  0  & 30 & 25  & 0 & 0   & 30  & 42 & 0  \\
III&  0  & 30 & 20  & 0 & 0   & 30  & 0  & 0  \\
IV & -30 & 30 & 25  & 0 & 30  & 30  & 0  & 0  \\
V  &  30 & 30 & 25  & 0 & -30 & 30  & 0  & 0  \\
\hline
\end{tabular}
\end{center}
\end{table}

Models I - V have no trend.  Models with trends will be considered below. The parameters for Model I were chosen to represent a scenario that is seasonally regular, with snow cover becoming present in the late Fall and staying until Spring ablation. The parameters $A_0$ and $A_0^*$ are set to zero, making the winter ``snow season" last roughly half the year. Model II has the same parameters as Model I, except that $\kappa^*$ was changed from 0 to 42, shifting the cosine wave governing $p_{1,0}(t)$ from its Model I settings. This change makes both $p_{0,1}(t)$ and $p_{1,0}(t)$ relatively large during the Spring months, which induces a Spring SCE season that oscillates more frequently between bare ground and snow cover.  Model III has the same parameters as Model I, except that $\kappa$ was changed from 25 to 20, making both $p_{0,1}(t)$ and $p_{1,0}(t)$ large during the Fall season.  This makes bare ground to snow cover oscillations more common in the Fall.   While we do not illustrate it here, increasing $A_1$ or $A_1^*$ tends to makes ``transitions" from winter to summer (and vice versa) shorter (sharper).  The parameters in Model IV are set to a lower latitude setting where snow only occurs sporadically during the middle of winter. This was done by decreasing the $A_0$ parameter from 0 to -30 for $p_{0,1}(t)$ and increasing $A_1^*$ from 0 to 30 (compared to Model I).  Model V's parameters correspond to a high latitude case where snow cover is present most of the year.  This was done by increasing $A_1$ from 0 to 30 and decreasing $A_1^*$ from 0 to -30 (compared to Model I).   These and other simulations show that the model can generate a wide range of SCE patterns.

\begin{figure}[ht]
\centering
\includegraphics[width=0.8\textwidth]{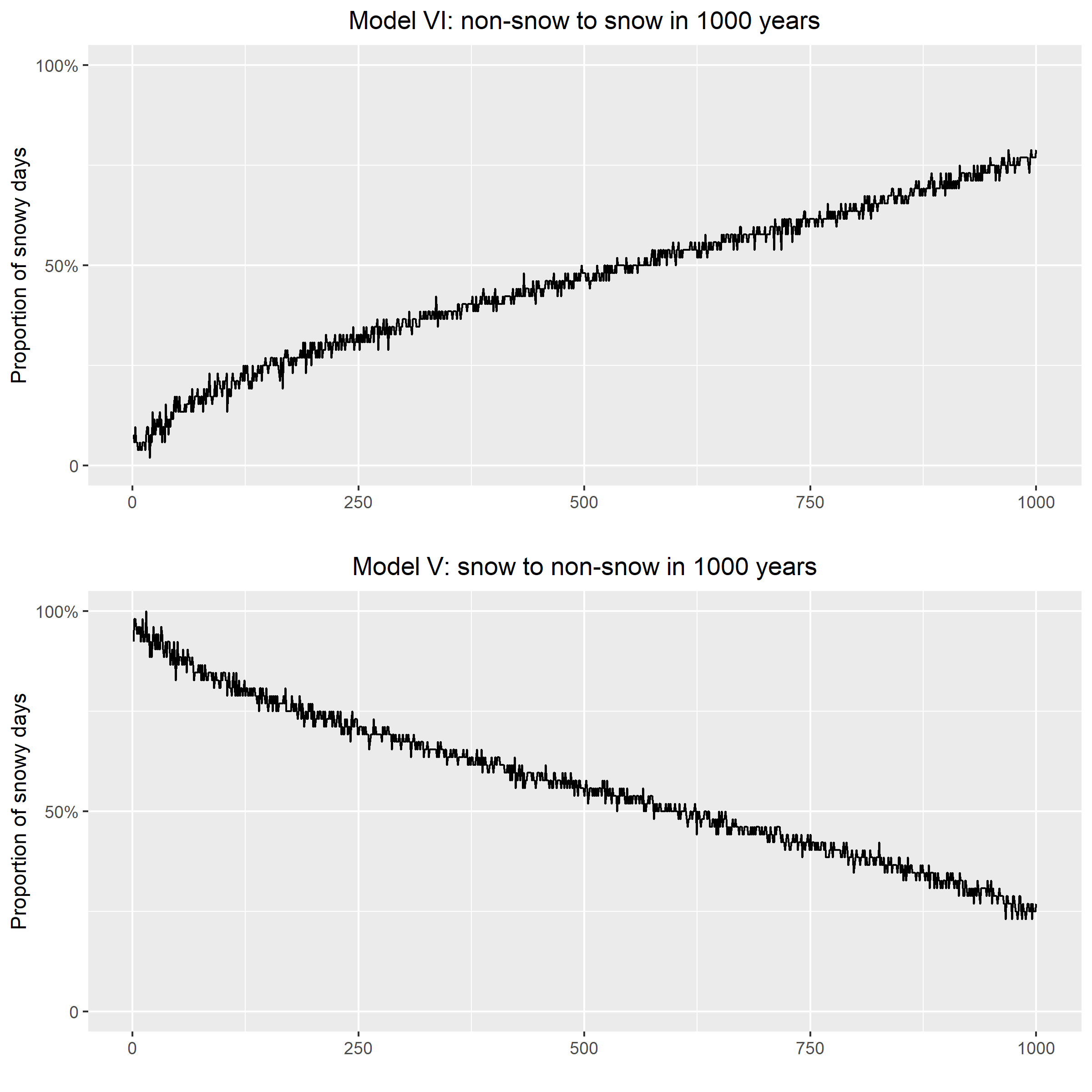}\\
\caption{Annual proportions of snowy days from Models IV and V with non-zero trends.}
\label{propor}
\end{figure}

To illustrate trend features, we choose parameters that bring Model IV above to a very snowy setting, and Model V above to a non-snowy scenario.  These are done over a 1000 year time period.  These scenarios are not climatologically realistic, but were chosen to demonstrate the overall flexibility of the approach. Figure \ref{propor} plots 
\[
\frac{1}{T} \sum_{\nu=1}^T X_{(k-1)T+\nu}
\]
against the annual index $k$. This quantity is the proportion of days of year $k$ where snow cover is present.  The top graphic in Figure \ref{sim} corresponds to Model IV, except that $\alpha$ was changed from zero to 0.001 and $\alpha^*$ is changed from zero to -0.001.  Here, the proportion of snow covered days rises from almost zero to approximately 80\%.  The antipodal scenario is illustrated in the bottom graphic of this figure. This moves a very snowy location to one without much snow cover.   This was done by taking Model V's parameters, but changing $\alpha$ from 0 to -0.001 and $\alpha^*$ from 0 to 0.001. 

\begin{figure}[ht]
\centering
\includegraphics[width=\textwidth]{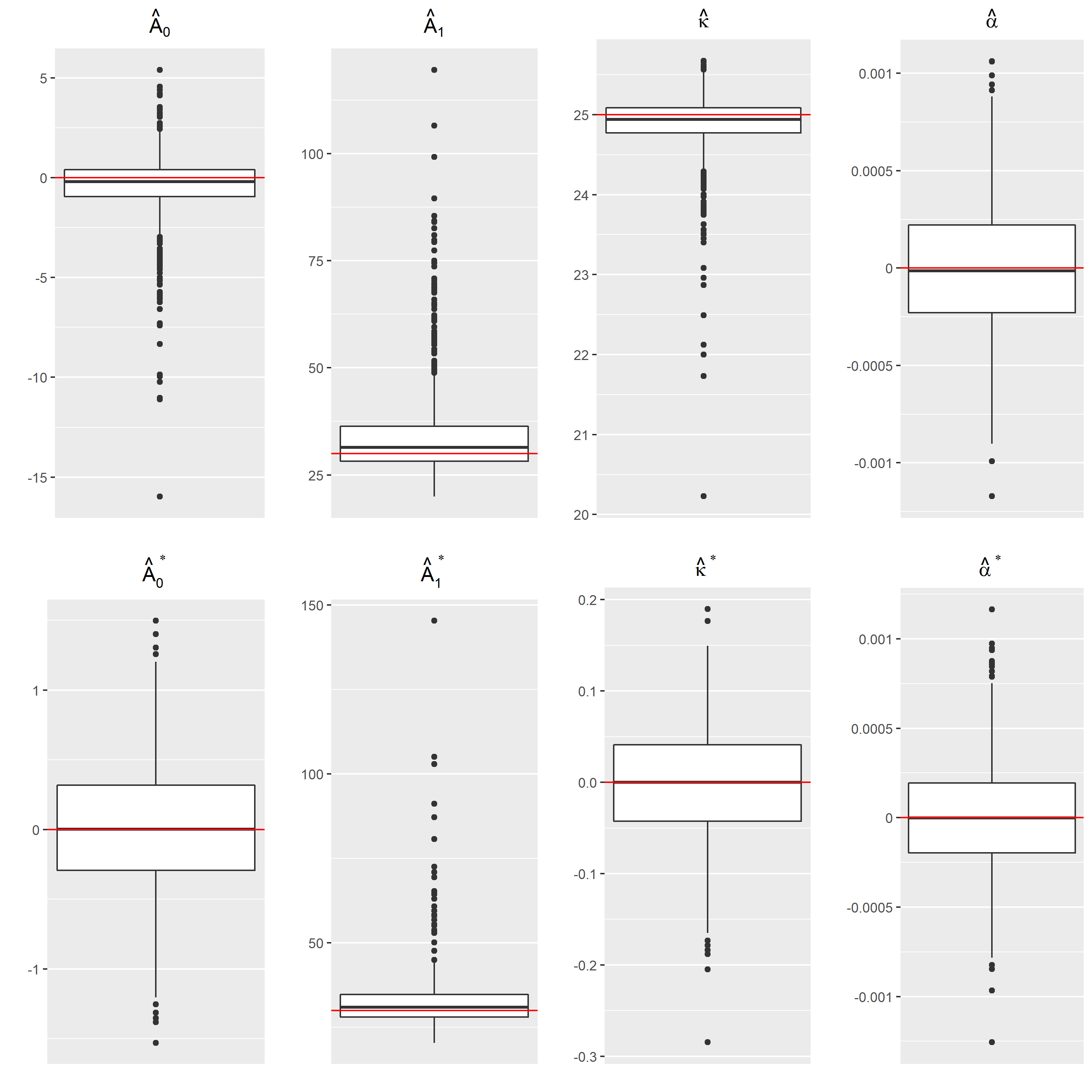} \\
\caption{Boxplots of the parameter estimates for each parameter from 1000 independent simulations. The red lines demarcate the true parameter values.}
\label{notrend}
\end{figure}

Turning to estimation, our first simulation case studies a 50-year series ($N=2600$), which is roughly the length of our satellite series studied later.   The parameters chosen for this simulation are those for Model I above; there is no trend in these simulations. These parameters were chosen to correspond to fitted parameters in some of our later analyzed grids.  Figure \ref{notrend} shows boxplots of the eight parameter estimators aggregated from 1000 independent simulations.  The solid line in each boxplot demarcates the median of the 1000 estimators for that parameter.  One sees little bias in the estimators.  Specifically, the estimation procedure was able to discern that there was no trend in the series.  Additional simulations (not shown here) indicate that any estimator bias recedes with increasing series length.  Estimation of the eight model parameters by likelihood appears to work well in this case.

\begin{figure}[ht]
\centering
\includegraphics[width=\textwidth]{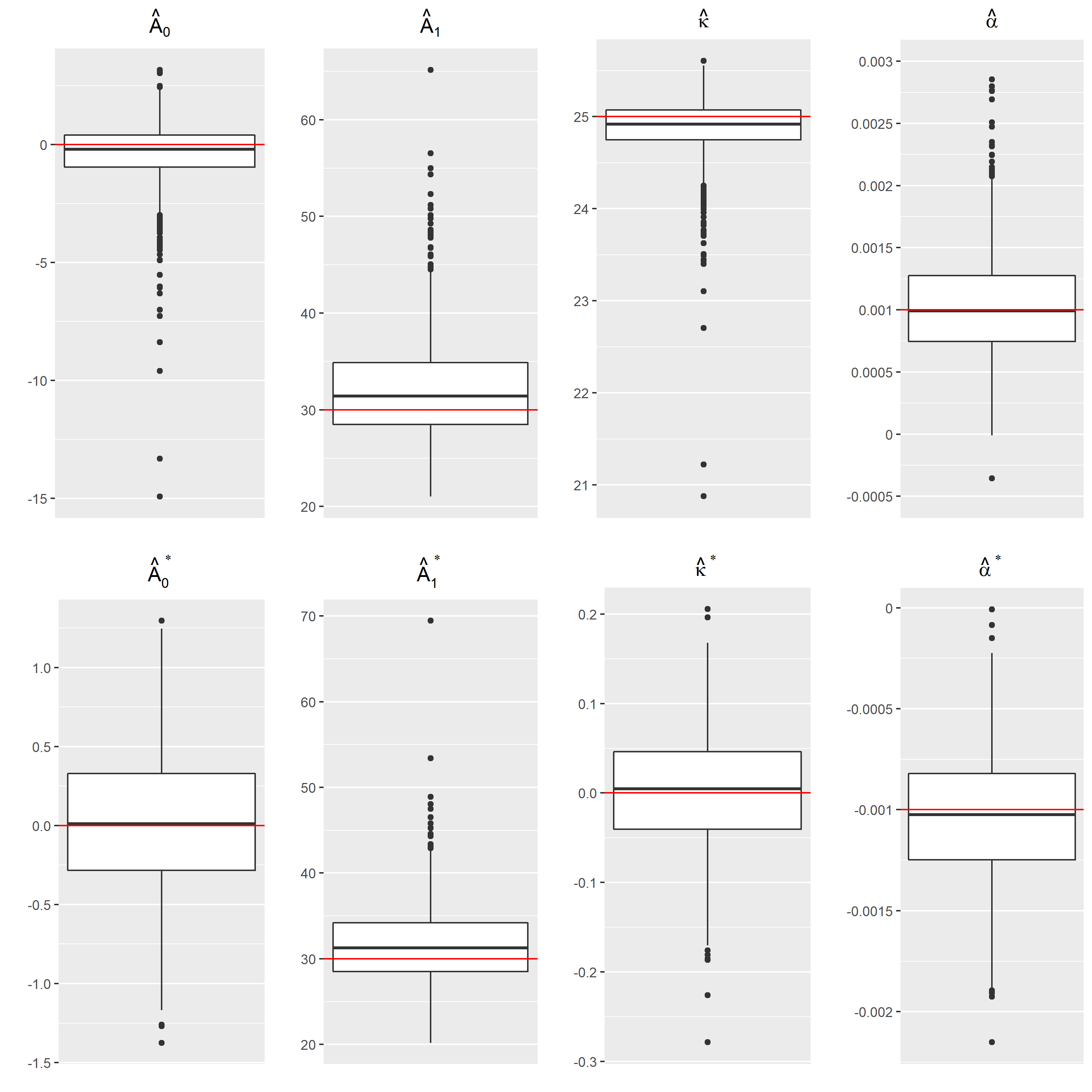}
\\
\caption{Boxplots of the parameter estimates for each parameter from 1000 independent simulations. The red lines demarcate the true parameter values.}
\label{no_changepoint}
\end{figure}

Our second simulation moves to a case with trends.   This simulation takes the same series length and parameters as the above simulation, but modifies the trend parameters to $\alpha =0.001$ and $\alpha ^*=-0.001$.  Figure \ref{no_changepoint} shows boxplots of the estimates of each parameter and are again quite good; importantly, trend parameters are accurately estimated.   While the trend parameters are small in magnitude in this simulation, they will be converted back to the scale of the problem (weeks of snow cover gained/lost per century) later for ease of interpretability.  Overall, model parameters are reasonably accurately estimated with 50 years of weekly satellite data.  

\section{A Sample Grid}
\label{SampleGrid}

\begin{figure}[ht]
\centering
\includegraphics[width=0.8\textwidth]{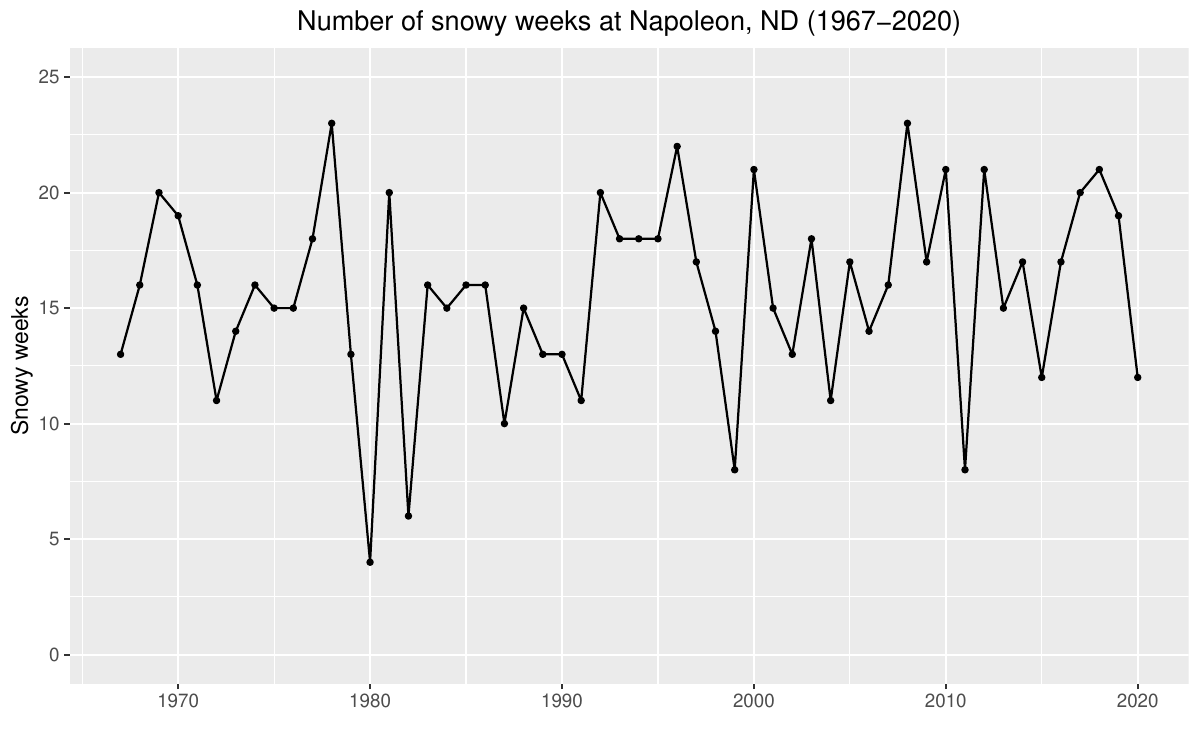}\\
\caption{Number of snow covered weeks during the 1967-2020 period for the Napoleon, ND grid.}
\label{sn}
\end{figure}

This section analyzes the snow coverage in the Napoleon, ND grid.  This grid covers a region which has been studied in previous snow studies \citep{Woody_WRR}.  Figure \ref{napoleon} displays a 10 year plot of the snow coverage at this grid and Figure \ref{sn} plots the number of snow covered weeks for each year from 1967-2020 at this grid.   This grid has never experienced more than 23 weeks of snow coverage in a winter (1978-79, 2008-09), nor less than 4 weeks (1980-81). A casual examination of this plot does not suggest declining snow presence --- this is scrutinized further below.

Table \ref{esti} below shows the maximum likelihood estimates of the parameters in the Section 3 model along with a single standard error.  All estimated parameters appear significantly non-zero except for the $\alpha$ parameters (one does not usually assess whether or not the phase shift parameters $\kappa$ and $\kappa^*$ are zero).  Statistical significance is assessed using asymptotic normality.  There is no statistical evidence to conclude that $\alpha$ is different from zero with a $p$-value of 0.7708, and decide that $p_{0,1}(t)$ is not changing.  As $p_{0,1}(t)$ governs transitions from bare ground to snow cover, this implies that the snow season is starting about the same time and has not changed over the study.  In contrast, $\alpha^*$ is concluded to be significantly negative with a $p$-value of 0.0001.   A negative $\alpha^*$ makes $p_{1,0}(t)$ smaller, which makes it harder for snow to disappear when it is on the ground.  This translates to a later Spring ablation.

\begin{table}[ht]
\caption{Model parameter estimates and their standard errors for the Napoleon, ND grid.}
\begin{center}
\begin{tabular}{|l|rrrr|}
\hline
Parameter  & $A_0$   & $A_1$   & $\kappa$ & $\alpha$ \\
           \hline
Estimate       & -3.2016 &  4.1499 & 24.3492  &0.0000382  \\
Standard Error &  0.2538 & 0.2936  & 0.26460  &0.0001315  \\
        \hline
        \hline
Parameter & $A_0^*$ & $A_1^*$ & $\kappa^*$ & $\alpha^*$ \\
        \hline
Estimate       &  1.7258 & 3.7889 & 49.8375 & -0.0004935 \\
Standard Error & 0.3774  & 0.4139 & 0.3800  &  0.0001273 \\
        \hline
    \end{tabular}
    \end{center}
    \label{esti}
\end{table}

To assess changes in snow presence, the $\hat{\beta}$ statistic in (\ref{trend}) is $\hat{\beta} = 0.038613$ and $\mbox{Var}(\hat{\beta})^{1/2} = 0.0247$. This translates to an additional 3.86 weeks of SCE over a century.  The test statistic for changing SCE is $Z = 1.5633$, which has a two-sided $p$-value of 0.1180.  This $p$-value is insignificant for a standard 5\% test, but is borderline significant for a 10\% test. Conclusions may change further if one sided alternative hypotheses are considered. The Napolean grid is experiencing increasing (and not decreasing) SCE changes.

\section{Results}

\begin{figure}[ht]
\centering
\includegraphics[width=1\textwidth]{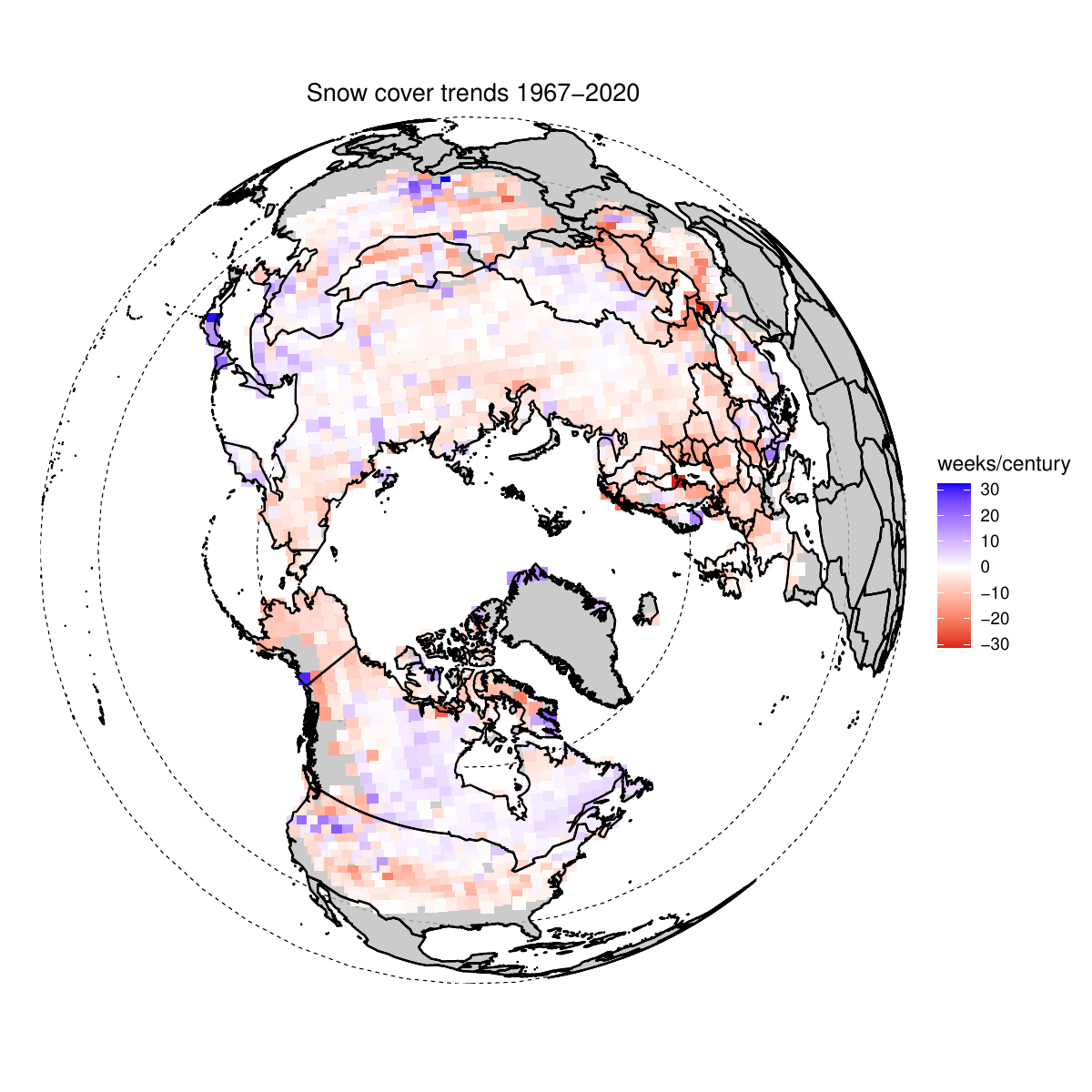}\\
\caption{Raw trends in the SCE data converted to weeks gained/lost per century.  Red and blue depict SCE losses and increases, respectively. Declining SCE grids outnumber advancing SCE grids by roughly a two to one ratio.}
\label{fig:trends}
\end{figure}

\begin{figure}[ht]
\centering
\includegraphics[width=1\textwidth]{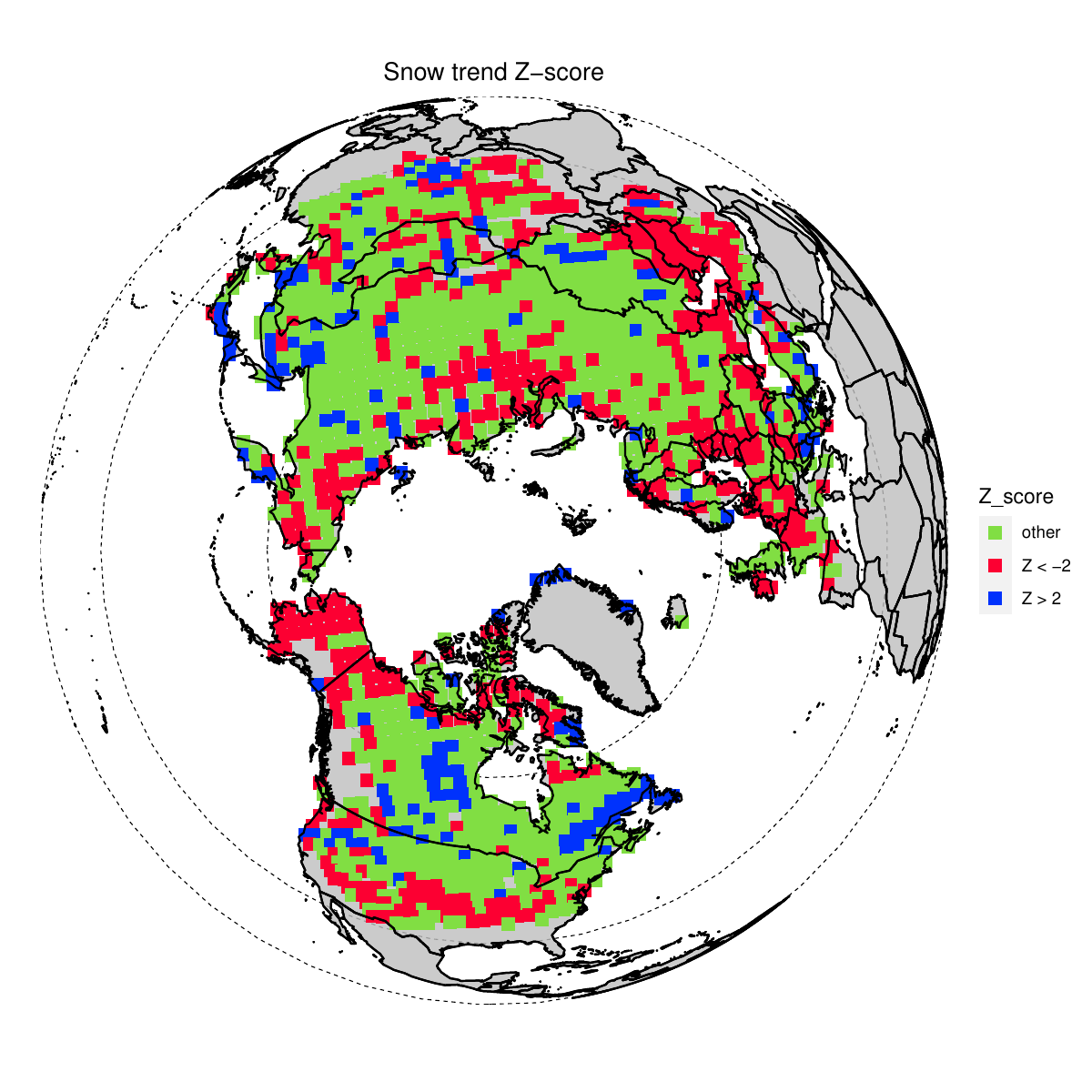}\\
\caption{$Z$ scores of the SCE trends. Trends in around half of the grids are not significantly non-zero.  Red indicates declining SCE and blue increasing SCE, with one-sided confidence at least 97.5\%.}
\label{fig:Z}
\end{figure}

This section reports results for the 1,613 grids where our model fit was deemed reliable.  Figure \ref{fig:trends} spatially portrays the trends $\hat{\beta}$ over all analyzed grids.  The corresponding $Z$-scores for the trend statistics are displayed in Figure \ref{fig:Z}.  In totality, 578 of the grids (35.83 \%) report a positive $\hat{\beta}$ (increasing snow), while 1035 grids (64.16 \%) show a negative $\hat{\beta}$.  This is almost a 2 to 1 margin preference for declining to advancing snow cover. The average trend over the 1,613 analyzed grids has lost 1.901 weeks of snow cover per century.

Examination of the spatial structure in Figures \ref{fig:trends} reveals regions of increasing and decreasing snow presence.  Decreasing snow presence in the Arctic, particularly in Russia and Western Canada and Alaska, is seen, agreeing with the findings of \cite{bormann2018estimating, estilow2015long}. Increasing snow is encountered in Eastern Canada, the Kamchatka Peninsula, and Japan.  Other regions experiencing positive trends can be seen in Figure \ref{fig:trends}.  The Figure \ref{fig:Z} $Z$-scores are deemed significantly non-zero should they exceed 2.0 in absolute value (the exact two-sided confidence is 0.9544).  Red colored $Z$-scores demarcate grids where snow cover is declining with at least 97.72\% confidence and blue colors depict increasing snow with 97.72\% confidence.  Overall, a general declining snow presence is seen along coastal areas and the periphery of the continental snowpack, with some inland increases in SCE, especially within North America.  This pattern could be associated with a deeper snowpack within continental interiors and a shallower or patchier snowpack along its edges, leading to more rapid retreat of the snowpack and a longer duration of its center. This coincides with the finding of the 4th IPCC report in \citep{lemke2007}.  

\begin{figure}[ht]
\centering
\includegraphics[width=1\textwidth]{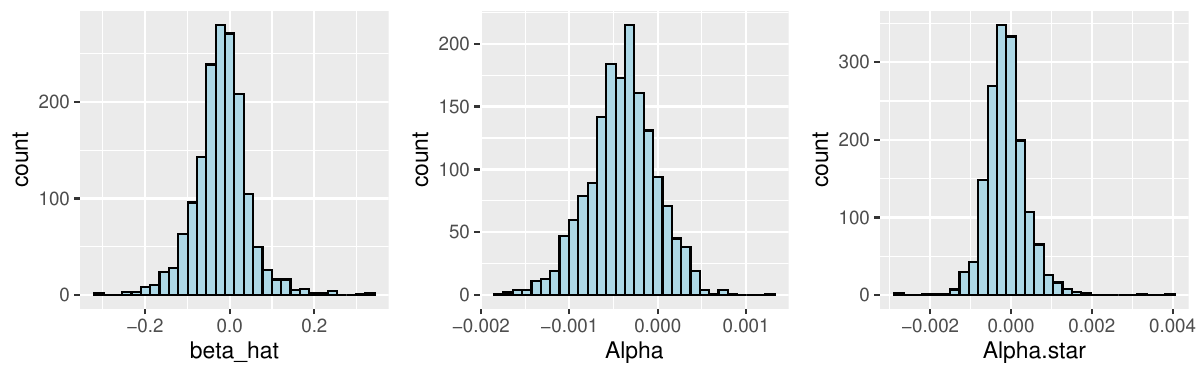}\\
\caption{Histograms over all 1,613 analyzed grids of (left) the estimated SCE trends $\hat{\beta}$, (center) the $\hat{\alpha}$ estimates, and (right) the $\hat{\alpha}^*$ estimates.  All histograms appear roughly unimodal (normal).  The mean of the left histogram is slightly negative.}
\label{fig:histo}
\end{figure}

The left panel in Figure \ref{fig:histo} shows a histogram of the trend estimates $\hat{\beta}$ over all analyzed grids. The estimated trends $\hat{\beta}$ are approximately normally distributed with a mean of -.01901 (the loss of 1.901 weeks of SCE per century alluded to above).  The center and right panels in Figure \ref{fig:histo} show histograms of the $\hat{\alpha}$ and $\hat{\alpha}^*$ parameters, respectively, over these same grids. The average $\alpha$ is -0.000398 and the average $\alpha^*$ is -0.000119. 


\begin{figure}[ht]
\centering
\includegraphics[width=1\textwidth]{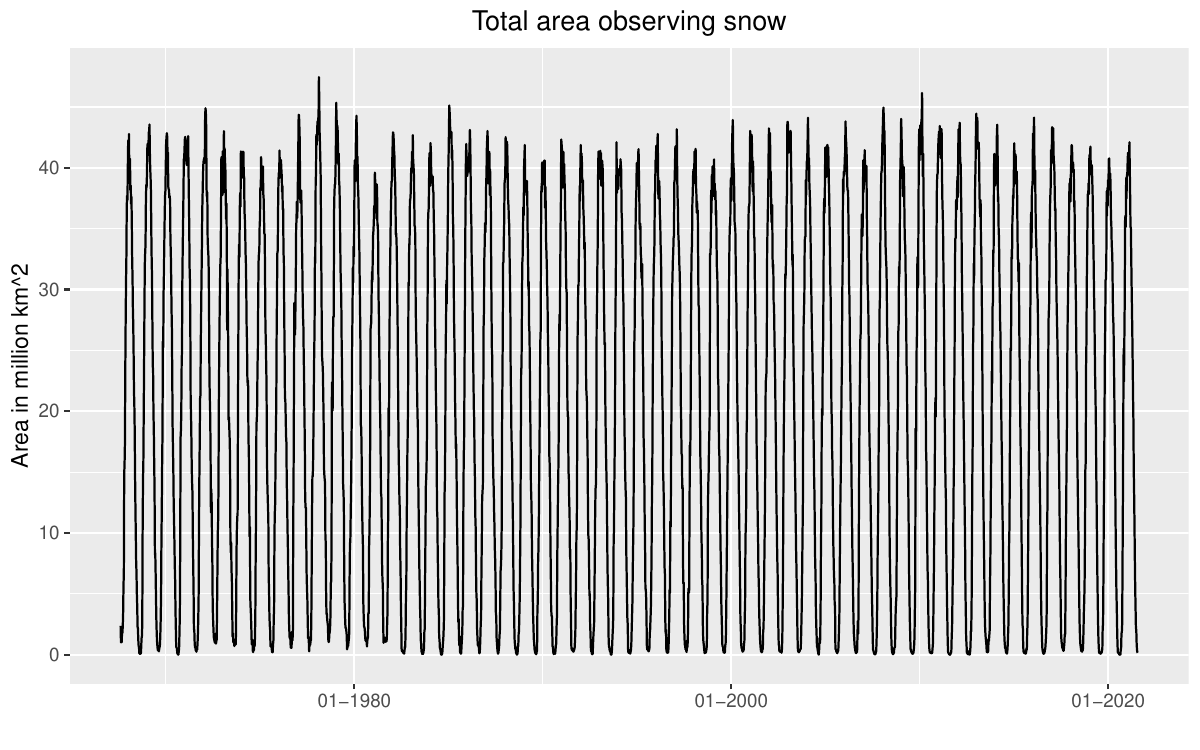} \\
\caption{Total SCE area by week over the period of record.   Trends are not visually obvious.}
\label{total}
\end{figure}

We now move to an investigation of temporal changes in the total SCE area.  Figure \ref{total} plots the total snow covered area in each week of the study over all analyzed grids.  Areas were obtained by adding the area of all snow covered grids; grid areas are included in the Rutgers Snow Lab SCE data and can be obtained from \url{http://climate.rutgers.edu/snowcover/}.  The seasonal cycle of SCE is evident, with winter weeks having the most prevalent snow cover.  While interannual variability is apparent, changes in this series are not visually evident in a visual inspection.

The Figure \ref{total} series is denoted by $\{ G_t \}$ and is now analyzed with a periodic linear regression. More on periodic regression analyses can be found in \cite{Lund_etal_1994} and \cite{Lund_2006}.  Our regression model for $G_t$ at time $t=nT+\nu$ is   
\begin{equation}
\label{periodic_SLR}
G_{nT+\nu} = \mu_\nu + \beta_\nu (nT+\nu) + \epsilon_{nT+\nu}.
\end{equation}
The parameter $\beta_\nu$ quantifies the linear rate of change in data during the $\nu$th week, for $1 \leq \nu \leq 52$; $\mu_\nu$ is a location parameter for week $\nu$.  The trend slope $\beta_\nu$ is allowed to depend on the week of year $\nu$, enabling us to investigate changes within a calendar year.  The regression errors $\{ \epsilon_t \}$ are assumed to have a zero mean for every week $\nu$.

The week $\nu$ trend $beta_\nu$ can be estimated via
\begin{equation}
\label{SLR_est}
\hat{\beta}_\nu =\frac{
\sum_{n=1}^d (G_{nT+\nu} - \bar{G}_\nu)(nT+\nu)}
{\sum_{n=1}^d (nT+\nu-\bar{t}_\nu)^2}
\end{equation}
\citep{Lund_etal_1994}.  Here, $\bar{t}_\nu= d^{-1}\sum_{n=1}^d (nT+\nu)=(d+1)T/2 +\nu$ and $\bar{G}_\nu= d^{-1}\sum_{n=1}^d G_{nT+\nu}$.  The denominator in (\ref{SLR_est}) can be worked out as $T^2d(d+1)(d-1)/12$.  We will not delve into standard error computations for $\hat{\beta}_\nu$, but refer the interested reader to \cite{Lund_etal_2001} for more on the issue.

\begin{figure}[ht]
\centering
\includegraphics[width=1\textwidth]{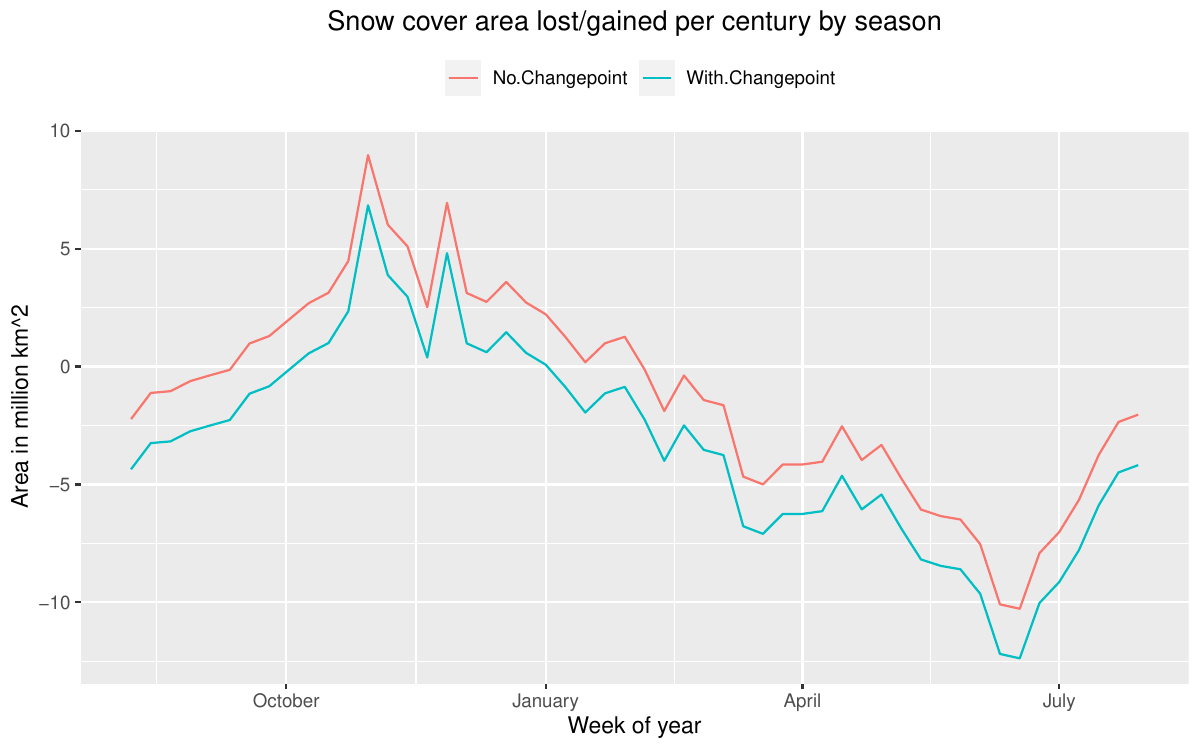} \\
\caption{Seasonal trend estimates SCE changes for each week of the year, scaled to area/gained lost per century. The estimated changepoint shift during the first week of June is $\hat{\Delta} = 0.794$ million km$^2$. Trends are uniformly lowered when a May 1999 changepoint for method changes to extract the SCE data is included in the model.}
\label{Seasonal_Trend_Changepoint}
\end{figure}

Figure \ref{Seasonal_Trend_Changepoint} plots estimates of $\beta_\nu$ against $\nu$ for each week of year (use the no changepoint graph).  Increasing SCE is evident in the Fall (late October through early December), with a corresponding decrease in late Winter through Summer. While increases span only a few months and include brief peaks above 5 million $km^2$, the decrease spans February - September, with losses below 5 million km$^2$ from May through July.  This implies that while the snow season is experiencing a shift toward an earlier onset and ablation period, there is a more pronounced decrease in snow cover through the warm season that is not being offset by increased snow in the Fall and early Winter.  Implications of this finding include a change in seasonal water availability.

As a final task of this section, we analyze possible issues induced by the methodological changes used to extract the snow cover data in May of 1999.  This will be done for the total SCE only; a deeper analysis exploring the effects on the individual grids is omitted.  To conduct this analysis, a shift in May of 1999 is allowed (this is also called an intervention since the changepoint time is known).  The setting is quantified through the regression
\begin{equation}
\label{periodic_SLR2}
G_{nT+\nu} = \mu_\nu + \beta_\nu (nT+\nu) + 
\Delta 1_{[nT+\nu \geq 1656]} + \epsilon_{nT+\nu}, \quad n=0, 1, \ldots, d-1.
\end{equation}
Here, 1656 is the week index of the May 1999 changepoint and $\Delta$ is its associated effect.  We do not see reason to allow $\Delta$ to depend on the week $\nu$, but could do so if desired.  

Estimates of the shift size are 
\begin{equation}
\label{SLR_est2}
\hat{\Delta}=0.794 ~\mbox{million km}^2.
\end{equation}
and a standard error for this estimate is 0.1432. Figure \ref{Seasonal_Trend_Changepoint} plots estimates of $\beta_\nu$ against $\nu$ for each week of year; see \cite{Lund_etal_1994} for the equations to fit this model. A $p$-value for the test that $\Delta=0$ is approximately $7.11 \times 10^{-9}$, indicating high confidence that the methodological change impacted observations, essentially making observations ''snowier".   In fact, the only positive trend slopes occur from October - December after the changepoint is taken into account.

\section{Summary and Comments}

This paper estimated Northern Hemispheric SCE trends over the last 55 years.  A flexible model was developed to quantify trends in periodic presence/absence data and assess their uncertainity margins.  Our SCE data were collected weekly and is count valued, taking the value of unity if snow is present and zero if snow cover is absent.  The data is periodic, with snow being more prevalent in the winter weeks. One contribution of this paper was to develop a model that adequately captures the data's periodicities and count structure. We were also able to assess the uncertainty margins of the trend estimates.  The developed model is highly flexible and could be fit to most grids in Europe, North America, and Asia that report snow.  In the most of the contiguous United States, trends could be reliably assessed down to latitudes of Prescott, AZ, Carlsbad, NM, and Knoxville, TN (the exception being some questionable SCE data from grids in mountainous area).

The results show that snow cover is declining overall, by a margin of almost 2 to 1 in terms of grid numbers. Arctic localities are showing heavy snow cover loss; however, other regions are experiencing increasing snow coverage, most notably Central and Eastern Canada and the Kamchatka and Japan vicinity.  Along with this general decline, a shift in the snow season towards an earlier onset and an earlier ablation period was seen, with the onset trending toward more snow in November and the ablation period showing declines from February through late Spring and early Summer.  The increased ablation in the warm season is not offset by the increased snow cover in the late Fall, possibly implying an overall change in the timing and distribution of water availability to regions that rely on spring snowmelt.

Statistical improvements can be made to this analysis. There is undoubtedly some non-zero spatial correlation in neighboring grids.  Accounting for spatial correlation would potentially lower uncertainty margins in the trend estimates; correlation usually does not appreciably change trend estimates, but accounting for correlation in multiple similar grids could reduce uncertainty margins in the trends.  Given the data quality issues present, the authors felt it more prudent to analyze the grids one by one and report which ones were ''unusable" (see two paragraphs below), which a spatial analysis would not do (at least initially).  It is also possible to smooth the Figure \ref{fig:trends} trends or their $Z$-scores in Figure \ref{fig:Z} in a spatial manner.  We did not pursue this here due to length concerns.

\begin{figure}[ht]
\centering
\includegraphics[width=1\textwidth]{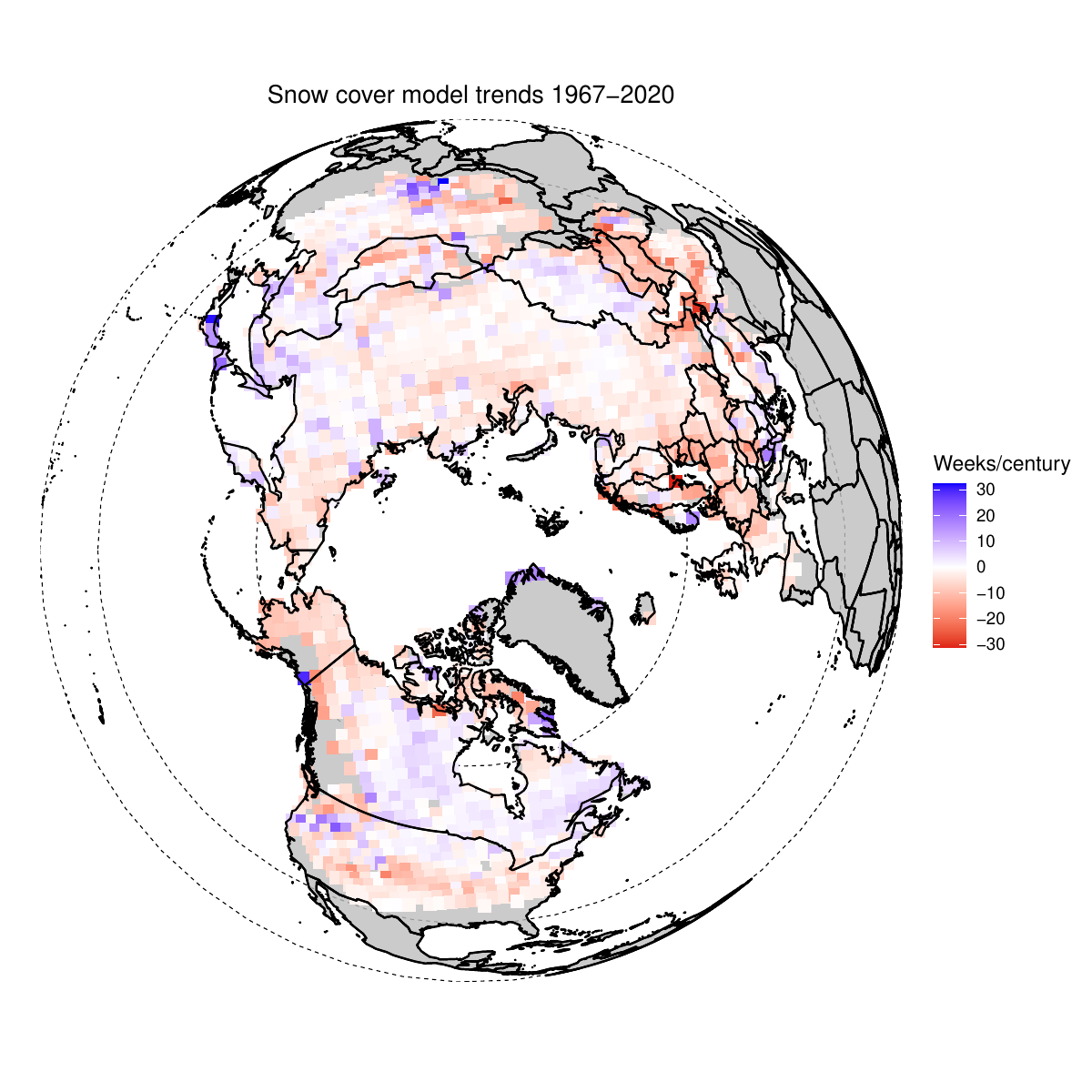} \\
\caption{Model-based trends estimated via (\ref{trend2}). The graphic looks similar to Figure \ref{fig:trends}.}
\label{fig:Model_Trend}
\end{figure}

The reader may note that our trend estimates are based on the data only and do not depend on the model (as it should be).  This said, one can also extract a trend estimate from the model.  One model-based trend is  
\begin{equation}
\label{trend2}
\frac{E[S_n] - E[S_1]}{n-1}.
\end{equation}
Both $E[S_n]$ and $E[S_1]$ are computed from the estimated model parameters, say computed ignoring the changepoint.  Figure \ref{fig:Model_Trend} shows a plot of these trends, converted to weeks of SCE gained/lost per century.  The graphic naturally resembles Figure \ref{fig:trends}.

Changepoints are discontinuity features in time series that occur at unknown times.  Changepoints often take place when measuring conditions change, such as station relocations or updates to gauge sensors.  Undoubtedly, some of the satellites or their related recording aspects changed during the period of record.  While we allowed for the known changepoint time in May of 1999 when the methods to extract the zero-one SCE data from the satellite pictures changed (this is a known changepoint time, which is also called an intervention or breakpoint), it would require significantly more work to find and adjust for these changepoints. Future work would assess grid changepoint features and homogenize the data from these grids. A caveat here:  while \cite{Lu_etal_2010} is a reference for changepoint methods for approximately normally distributed temperature data, methods to homogenize zero-one count data have have yet to be developed (or have not matured) in the statistics literature.

While most of the grids here report what appears to be high quality data, the green-colored grids in Figure \ref{fig:group} flag grids reporting suspect data.  We hope that these grids can be reexamined/fixed in the future for inclusion in studies such as this.

\clearpage
\acknowledgments

Robert Lund, Yisu Jia, Jiajie Kong, and Jon Woody thank NSF Grant DMS 2113592 for partial support.

%
%
\datastatement The Northern Hemispheric snow cover extend data in this study are available at Rutgers University Snow Lab available at \url{http://climate.rutgers.edu/snowcover/}. We used the data from August 1967 to July 2021 for 54 winter-centered years.


%

\appendix
We start by computing $\mbox{Cov}(S_n, S_{n+h})$ for every $h > 0$ and $n$ in $\{ 1, \ldots, d-h \}$.  For this, $\mbox{Cov}(S_n,S_{n+h}) = E[S_nS_{n+h}]- E[S_n]E[S_{n+h}]$. To get $E[ S_n ]$, use
\[
E[S_n] = E\left[\sum_{\nu=1}^T 1_{[X_{(n-1)T+\nu} = 1]} \right]
       =        \sum_{\nu=1}^T P[ X_{(n-1)T+\nu}=1]
       =        \sum_{\nu=1}^T \pi_1((n-1)T+\nu).
\]
This quantity will need to be estimated/evaluated at the model's maximum likelihood parameters.

The calculation of $E[S_nS_{n+h}]$ for $h>0$ is a little more delicate.  First, suppose that $h > 0$; the case where $h = 0$ will be handled separately. Then
\begin{align*}
E[S_nS_{n+h}] & = E \left[ \left( 
\sum_{u=1}^T 1_{[X_{(n-1)T+u}= 1)]} \right)
\left( \sum_{\nu=1}^T1_{[X_{(n+h-1)T+\nu}=1]} \right)\right]\\
&= \sum_{u=1}^{T} \sum_{\nu=1}^{T} 
P[ X_{(n-1)T+u}= 1 \cap X_{(n+h-1)T+\nu}=1 ] \\    
&=\sum_{u=1}^{T}\sum_{\nu=1}^{T} 
P[ X_{(n-1)T+u}= 1] P[ X_{(n+h-1)T+\nu}= 1 | X_{(n-1)T+u}= 1] \\
&= \sum_{u=1}^{T} \sum_{\nu=1}^{T} 
\pi_1(t_1) {\bf P}^*(t_1,t_2)_{2,2},   \\    
\end{align*}
where, $t_1= (n-1)T_+u$, $t_2=(n-1)T+\nu$, and the notation ${\bf A}_{i,j}$ denotes the element in the $i$th row of the $j$th column of the matrix ${\bf A}$.

For the case where $h = 0$, direct computation yields
\begin{align*}
E[S_n^2] 
& =  E \left[ \sum_{u=1}^T (1_{[X_{(n-1)T+u}=1]})^2 \right] + 
2 E \left[ \sum_{u=1}^T \sum_{v=u+1}^T 
1_{[X_{(n-1)T+u}=1]} \times 1_{[X_{(n-1)T+v}= 1]} \right] \\
& = \sum_{u=1}^T P[ X_{(n-1)T+u}=1] +
  2 \sum_{u=1}^T \sum_{v=u+1}^T
  P[ X_{(n-1)T+u}=1 \cap X_{(n-1)T+v} = 1) ]\\
&= \sum_{u=1}^T \pi_1(t_1) +
  2 \sum_{u=1}^T \sum_{v = u+1}^T \pi_1(t_1) {\bf P}(t_1,t_2)_{2,2}
\end{align*}
after the relation $1_A^2 = 1_A$ is applied.  Here, $t_1=(n-1)T+u$ and 
$t_2= (n-1)T+\nu$.  This calculation allows us to compute $\mbox{Cov}(S_i,S_j)$ for every $i$ and $j$.





%



\bibliographystyle{ametsocV6}
\bibliography{SnowTrends}

\begin{thebibliography}{28}
\providecommand{\natexlab}[1]{#1}
\providecommand{\url}[1]{\texttt{#1}}
\renewcommand{\UrlFont}{\rmfamily}
\providecommand{\urlprefix}{URL }
\expandafter\ifx\csname urlstyle\endcsname\relax
  \providecommand{\doi}[1]{https://doi.org/\discretionary{}{}{}#1}\else
  \providecommand{\doi}{https://doi.org/\discretionary{}{}{}\begingroup
  \urlstyle{rm}\Url}\fi
\providecommand{\eprint}[2][]{\url{#2}}

\bibitem[{Barnett et~al.(2005)Barnett, Adam,, and
  Lettenmaier}]{barnett2005potential}
Barnett, T.~P., J.~C. Adam, and D.~P. Lettenmaier, 2005: Potential impacts of a
  warming climate on water availability in snow-dominated regions.
  \textit{Nature}, \textbf{438~(7066)}, 303--309.

\bibitem[{Bormann et~al.(2018)Bormann, Brown, Derksen,, and
  Painter}]{bormann2018estimating}
Bormann, K.~J., R.~D. Brown, C.~Derksen, and T.~H. Painter, 2018: Estimating
  snow-cover trends from space. \textit{Nature Climate Change},
  \textbf{8~(11)}, 924--928.

\bibitem[{Brown et~al.(2007)Brown, Derksen,, and Wang}]{brown2007assessment}
Brown, R., C.~Derksen, and L.~Wang, 2007: Assessment of spring snow cover
  duration variability over northern {C}anada from satellite datasets.
  \textit{Remote Sensing of Environment}, \textbf{111~(2-3)}, 367--381.

\bibitem[{Brown and Robinson(2011)Brown, and Robinson}]{brown2011northern}
Brown, R.~D., and D.~A. Robinson, 2011: {N}orthern {H}emisphere spring snow
  cover variability and change over 1922--2010 including an assessment of
  uncertainty. \textit{The Cryosphere}, \textbf{5~(1)}, 219--229.

\bibitem[{Callaghan et~al.(2011)}]{callaghan2011changing}
Callaghan, T.~V., and Coauthors, 2011: The changing face of {A}rctic snow
  cover: {A} synthesis of observed and projected changes. \textit{Ambio},
  \textbf{40~(1)}, 17--31.

\bibitem[{Chib and Greenberg(1998)Chib, and Greenberg}]{Probit}
Chib, S., and E.~Greenberg, 1998: Analysis of multivariate probit models.
  \textit{Biometrika}, \textbf{85~(2)}, 347--361.

\bibitem[{D{\'e}ry and Brown(2007)D{\'e}ry, and Brown}]{dery2007recent}
D{\'e}ry, S.~J., and R.~D. Brown, 2007: Recent {N}orthern {H}emisphere snow
  cover extent trends and implications for the snow-albedo feedback.
  \textit{Geophysical Research Letters}, \textbf{34~(22)},
  \doi{10.1029/2007GL031474}.

\bibitem[{Dye(2002)}]{dye2002variability}
Dye, D.~G., 2002: Variability and trends in the annual snow-cover cycle in
  {N}orthern {H}emisphere land areas, 1972--2000. \textit{Hydrological
  Processes}, \textbf{16~(15)}, 3065--3077.

\bibitem[{Estilow et~al.(2015)Estilow, Young,, and Robinson}]{estilow2015long}
Estilow, T.~W., A.~H. Young, and D.~A. Robinson, 2015: A long-term {N}orthern
  {H}emisphere snow cover extent data record for climate studies and
  monitoring. \textit{Earth System Science Data}, \textbf{7~(1)}, 137--142.

\bibitem[{Goudie(2018)}]{goudie2018human}
Goudie, A.~S., 2018: \textit{Human {I}mpact on the {N}atural {E}nvironment}.
  John Wiley \& Sons, Hoboken, NJ, USA.

\bibitem[{Karl et~al.(2009)Karl, Melillo, Peterson,, and
  Hassol}]{karl2009global}
Karl, T.~R., J.~M. Melillo, T.~C. Peterson, and S.~J. Hassol, 2009:
  \textit{Global {C}limate {C}hange {I}mpacts in the {U}nited {S}tates}.
  Cambridge University Press, Cambridge, United Kingdom and New York, NY, USA.

\bibitem[{Lawrence and Slater(2010)Lawrence, and
  Slater}]{lawrence2010contribution}
Lawrence, D.~M., and A.~G. Slater, 2010: The contribution of snow condition
  trends to future ground climate. \textit{Climate Dynamics}, \textbf{34~(7)},
  969--981.

\bibitem[{Lemke et~al.(2007)}]{lemke2007}
Lemke, P., and Coauthors, 2007: {O}bservations: {C}hanges in {S}now, {I}ce, and
  {F}rozen {G}round. \textit{Climate Change 2007: The Physical Science Basis.
  Contribution of Working Group I to the Fourth Assessment Report of the
  Intergovenmental Panel on Climate Change}, S.~Solomon et~al., Eds., Cambridge
  University Press, Cambridge, United Kingdom and New York, NY, USA.

\bibitem[{Liston and Hiemstra(2011)Liston, and Hiemstra}]{liston2011changing}
Liston, G.~E., and C.~A. Hiemstra, 2011: The changing cryosphere:
  {P}an-{A}rctic snow trends (1979--2009). \textit{Journal of Climate},
  \textbf{24~(21)}, 5691--5712.

\bibitem[{Lu et~al.(2010)Lu, Lund,, and Lee}]{Lu_etal_2010}
Lu, Q., R.~B. Lund, and T.~C.~M. Lee, 2010: An {MDL} approach to the climate
  segmentation problem. \textit{Annals of Applied Statistics}, \textbf{4},
  299--319.

\bibitem[{Lund(2006)}]{Lund_2006}
Lund, R., 2006: A seasonal analysis of riverflow trends. \textit{Journal of
  Statistical Computation and Simulation}, \textbf{76}, 397--405.

\bibitem[{Lund et~al.(1995)Lund, Hurd, Bloomfield,, and Smith}]{Lund_etal_1994}
Lund, R., H.~Hurd, P.~Bloomfield, and R.~Smith, 1995: Climatological time
  series with periodic correlation. \textit{Journal of Climate},
  \textbf{8~(11)}, 2787--2809.

\bibitem[{Lund et~al.(2001)Lund, Seymour,, and Kafadar}]{Lund_etal_2001}
Lund, R., L.~Seymour, and K.~Kafadar, 2001: Temperature trends in the {U}nited
  {S}tates. \textit{Environmetrics}, \textbf{12}, 673--690.

\bibitem[{Mote(2003)}]{mote2003trends}
Mote, P.~W., 2003: Trends in snow water equivalent in the {P}acific {N}orthwest
  and their climatic causes. \textit{Geophysical Research Letters},
  \textbf{30~(12)}, \doi{10.1029/2003GL017258}.

\bibitem[{Mote et~al.(2018)Mote, Li, Lettenmaier, Xiao,, and
  Engel}]{mote2018dramatic}
Mote, P.~W., S.~Li, D.~P. Lettenmaier, M.~Xiao, and R.~Engel, 2018: Dramatic
  declines in snowpack in the western {U.S.} \textit{Npj Climate and
  Atmospheric Science}, \textbf{1~(1)}, 1--6.

\bibitem[{Notarnicola(2022)}]{notarnicola2022overall}
Notarnicola, C., 2022: Overall negative trends for snow cover extent and
  duration in global mountain regions over 1982--2020. \textit{Scientific
  Reports}, \textbf{12~(1)}, 1--16.

\bibitem[{Robinson et~al.(1993)Robinson, Dewey,, and Heim}]{robinson1993global}
Robinson, D.~A., K.~F. Dewey, and R.~R. Heim, 1993: Global snow cover
  monitoring: {A}n update. \textit{Bulletin of the American Meteorological
  Society}, \textbf{74~(9)}, 1689--1696.

\bibitem[{Serreze et~al.(2000)}]{serreze2000observational}
Serreze, M., and Coauthors, 2000: Observational evidence of recent change in
  the northern high-latitude environment. \textit{Climatic Change},
  \textbf{46~(1)}, 159--207.

\bibitem[{Van~Mantgem et~al.(2009)}]{van2009widespread}
Van~Mantgem, P.~J., and Coauthors, 2009: Widespread increase of tree mortality
  rates in the western {U}nited {S}tates. \textit{Science},
  \textbf{323~(5913)}, 521--524.

\bibitem[{Wiesnet et~al.(1987)Wiesnet, Ropelewski, Kukla,, and
  Robinson}]{wiesnet1987discussion}
Wiesnet, D., C.~Ropelewski, G.~Kukla, and D.~Robinson, 1987: A discussion of
  the accuracy of {NOAA} satellite-derived global seasonal snow cover
  measurements. \textit{Large Scale Effects of Seasonal Snow Cover},
  \textbf{166}, 291--304.

\bibitem[{Woody et~al.(2009)Woody, Lund, Grundstein,, and Mote}]{Woody_WRR}
Woody, J., R.~Lund, A.~J. Grundstein, and T.~L. Mote, 2009: A storage model
  approach to the assessment of snow depth trends. \textit{Water Resources
  Research}, \textbf{45~(10)}, \doi{10.1029/2009WR007996}.

\bibitem[{Yue et~al.(2002)Yue, Pilon, Phinney,, and
  Cavadias}]{yue2002influence}
Yue, S., P.~Pilon, B.~Phinney, and G.~Cavadias, 2002: The influence of
  autocorrelation on the ability to detect trend in hydrological series.
  \textit{Hydrological Processes}, \textbf{16~(9)}, 1807--1829.

\bibitem[{Zona et~al.(2016)}]{zona2016cold}
Zona, D., and Coauthors, 2016: Cold season emissions dominate the {A}rctic
  tundra methane budget. \textit{Proceedings of the National Academy of
  Sciences}, \textbf{113~(1)}, 40--45.

\end{thebibliography}

\end{document}